\documentclass[preprint2]{proto}
\usepackage{times}

\newcommand{\refs}{\par\noindent\hangindent=1pc\hangafter=1}
\newcommand\msun{M_{\odot}}
\newcommand\msunyr{M_{\odot}\,\rm yr^{-1}}
\newcommand\mdot{\dot{M}}

\begin{document}

\title{\textbf{\LARGE The Formation of Brown Dwarfs: Observations}}

\author {\textbf{\large Kevin L. Luhman}}
\affil{\small\em The Pennsylvania State University}
\author {\textbf{\large Viki Joergens}}
\affil{\small\em  The University of Leiden}
\author {\textbf{\large Charles Lada}}
\affil{\small\em Smithsonian Astrophysical Observatory}
\author {\textbf{\large James Muzerolle and Ilaria Pascucci}}
\affil{\small\em The University of Arizona}
\author {\textbf{\large Russel White}}
\affil{\small\em The University of Alabama}

\begin{abstract}
\baselineskip = 11pt
\leftskip = 0.65in
\rightskip = 0.65in
\parindent=1pc
{\small
We review the current state of observational work on the formation of 
brown dwarfs, focusing on their initial mass function, velocity and spatial
distributions at birth, multiplicity, accretion, and circumstellar disks.
The available measurements of these various properties are consistent with 
a common formation mechanism for brown dwarfs and stars. 
In particular, the existence of widely separated binary brown dwarfs and 
a probable isolated proto-brown dwarf indicate that some substellar objects 
are able to form in the same manner as stars through unperturbed cloud 
fragmentation. Additional mechanisms such as ejection and photoevaporation 
may play a role in the birth of some brown dwarfs, but 
there is no observational evidence to date to suggest that they are 
the key elements that make it possible for substellar bodies to form.
 \\~\\~\\~}

\end{abstract}  

\section{\textbf{INTRODUCTION}}

Although many of the details are not perfectly understood,
stars and giant planets are generally believed to form through the collapse of 
molecular cloud cores and the accretion of gas by rocky cores in 
circumstellar disks, respectively.
In comparison, the formation of objects intermediate between stars and
planets -- free-floating and companion brown dwarfs -- has no widely
accepted explanation. A priori, one might expect that brown dwarfs form in
the same manner as stars, just on a much smaller scale.
However, although self-gravitating objects can form with initial masses of only 
$\sim1$~$M_{\rm Jup}$ in simulations of the fragmentation of
molecular cloud cores, these fragments continue to accrete matter from 
their surrounding cores, usually to the point of eventually reaching 
stellar masses ({\em Boss}, 2001; {\em Bate et al.}, 2003).
Thus, standard cloud fragmentation in these models seems to have difficulty
in making brown dwarfs. 
One possible explanation is that the simulations lack an important
piece of physics (e.g., turbulence), and brown dwarfs are able to form
through cloud fragmentation
despite their predictions (e.g., {\em Padoan and Nordlund}, 2004). 
Another possibility is that a brown dwarf is
born when cloud fragmentation is modified by an additional process that 
prematurely halts accretion during the protostellar stage, 
such as dynamical ejection 
({\em Reipurth and Clarke}, 2001; {\it Boss}, 2001; {\em Bate et al.}, 2002)
or photoevaporation by ionizing radiation from massive stars 
({\em Kroupa and Bouvier}, 2003; {\em Whitworth and Zinnecker}, 2004).
This uncertainty surrounding the formation of brown dwarfs has motivated a
great deal of theoretical and observational work over the last decade.

\begin{figure*}
\epsscale{2.2}
\plottwo{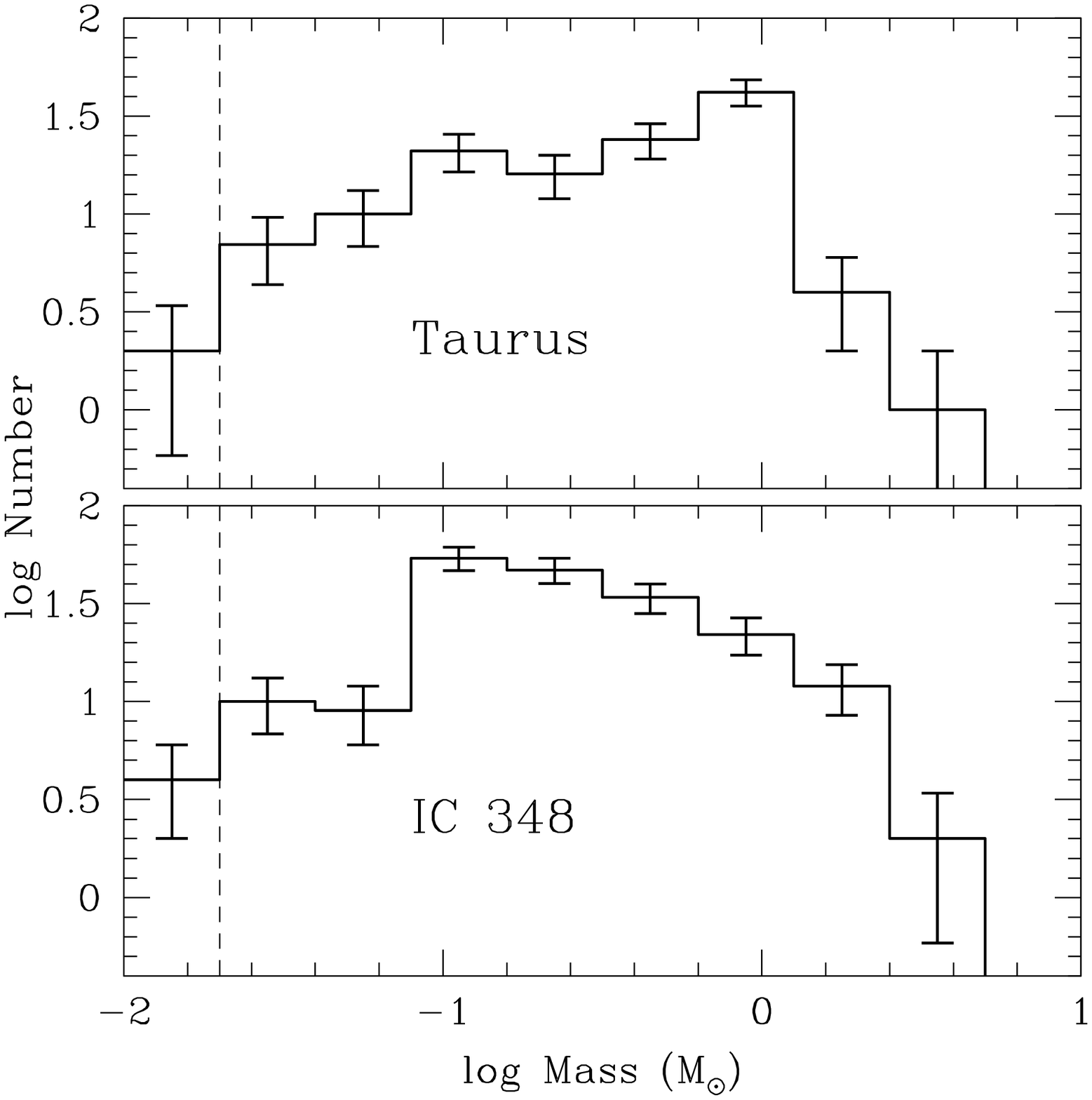}{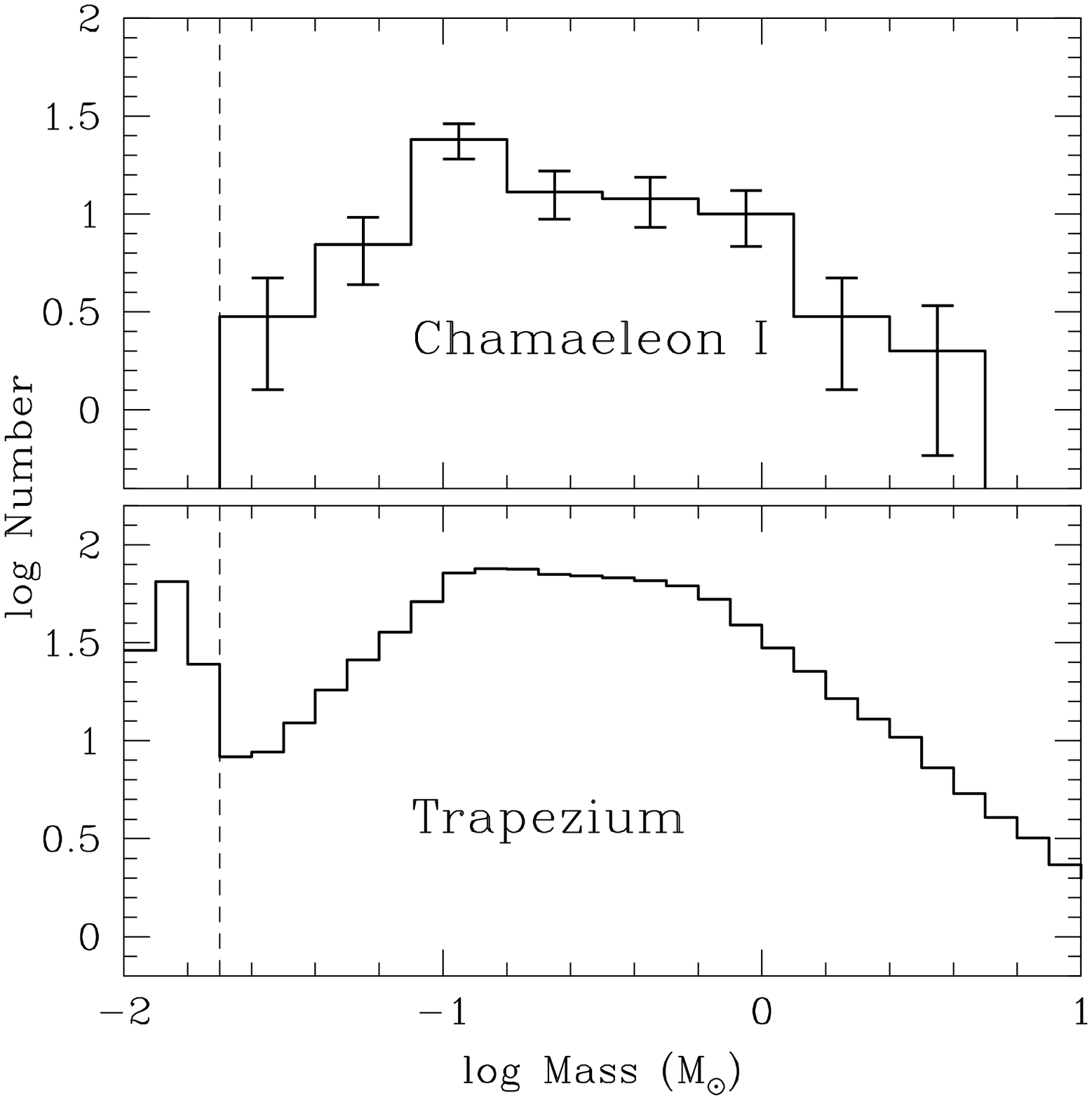}
\caption{\small 
IMFs for Taurus ({\em Luhman}, 2004c), IC~348 ({\em Luhman et al.}, 2003b),
Chamaeleon I ({\em Luhman}, in preparation), and the Trapezium Cluster
({\em Muench et al.}, 2002). The completeness limits for these measurements
are near 0.02~$M_{\odot}$ ({\it dashed lines}). 
In the units of this diagram, the Salpeter slope is 1.35.
\label{fig:imf}}
\end{figure*}

In this paper, we review the current observational constraints on the 
formation process of brown dwarfs (BDs), which complements the theoretical
review of this topic provided in the chapter by {\em Whitworth et al.}
By the nature of the topic of this review, we focus on 
observations of BDs at young ages ($\tau<10$~Myr), although we also 
consider properties of evolved BDs that provide insight into BD formation
(e.g., multiplicity). A convenient characteristic of young BDs is their 
relatively bright luminosities and warm temperatures compared to their
older counterparts in the solar neighborhood, making them easier to observe.
However, because the luminosities and temperatures of young BDs
are continuous extensions of those of stars, positively identifying 
a young object as either a low-mass star or a BD is often not 
possible. The mass estimates for a given object vary greatly with
the adopted evolutionary models and the manner in which observations are
compared to the model predictions. Using the models of 
{\em Baraffe et al.} (1998) and {\em Chabrier et al.} (2000) 
and the temperature scale of {\em Luhman et al.} (2003b),
the hydrogen burning mass limit at ages of 0.5-3~Myr corresponds to a spectral
type of $\sim$M6.25, which is consistent with the dynamical mass and
spectral type of the first known eclipsing binary BD
({\em Stassun et al.}, 2006) and other observational tests
({\em Luhman and Potter}, 2006). Therefore, we will treat young objects 
later than M6 as BDs for the purposes of this review.

\section{\textbf{INITIAL MASS FUNCTION}}
\label{sec:imf}

One of the most fundamental properties of BDs is their initial mass function
(IMF). Because BDs are brightest when they are young, star-forming regions and
young clusters are the best sites for finding them in large numbers
and at low masses, which is necessary for measuring statistically significant
IMFs. Spectroscopic surveys for BDs have been performed toward many young 
populations ($\tau<10$~Myr) during the last decade, including
IC~348 ({\em Luhman et al.}, 1998, 2003b, 2005a; {\em Luhman}, 1999),
Taurus ({\em Brice\~no et al.}, 1998, 2002;
{\em Mart{\'\i}n et al.}, 2001b; 
{\em Luhman}, 2000, 2004c, 2006; {\em Luhman et al.}, 2003a; 
{\em Guieu et al.}, 2006),
Chamaeleon~I 
({\em Com\'eron et al.}, 1999, 2000, 2004; 
{\em Neuh\"{a}user and Comer\'{o}n}, 1999;
{\em Luhman}, 2004a,b; {\em Luhman et al.}, 2004),
Ophiuchus ({\em Luhman et al.}, 1997; {\em Wilking et al.}, 1999; 
{\em Cushing et al.}, 2000),
Upper Scorpius ({\em Ardila et al.}, 2000; {\em Mart{\'\i}n et al.}, 2004),
Orion ({\em Hillenbrand}, 1997; {\em Lucas et al.}, 2001;
{\em Slesnick et al.}, 2004),
NGC~2024 ({\em Levine et al.}, 2006),
NGC~1333 ({\em Wilking et al.}, 2004),
TW~Hya ({\em Gizis}, 2002; {\em Scholz et al.}, 2005),
$\lambda$~Ori ({\em Barrado y Navascu\'es et al.}, 2004b),
and $\sigma$ Ori 
({\em Barrado y Navascu\'es et al.}, 2001, 2002; 
{\em B\'ejar et al.}, 1999, 2001; 
{\em Mart{\'\i}n et al.}, 2001a; 
{\em Zapatero Osorio et al.}, 1999, 2000, 2002a,b,c; 
{\em Mart{\'\i}n and Zapatero Osorio}, 2003).

We now examine the IMF measurements for IC~348, Chamaeleon~I, Taurus, and
the Trapezium, which exhibit the best combination of number statistics,
completeness, and dynamic range in mass among the young populations studied 
to date. These IMFs are shown in Fig.~\ref{fig:imf}.
Because the same techniques and models were employed in converting
from data to masses for each population, one can be confident in the validity
of any differences in these IMFs.
For the Trapezium, we use the IMF derived through infrared (IR) 
luminosity function modeling by {\em Muench et al.} (2002)
(see also {\em Luhman et al.}, 2000; {\em Hillenbrand and Carpenter}, 2000;  
{\em Lucas et al.}, 2005).
The spectroscopically determined IMF for IC~348 agrees well with the IMF derived
by {\em Muench et al.} (2003) 
through the same kind of luminosity function analysis,
which suggests that the Trapezium IMF from {\em Muench et al.} (2002)
can be reliably compared to the spectroscopic IMFs for IC~348, 
Chamaeleon~I, and Taurus. We quantify the relative numbers of
BDs and stars with the ratio:
 
$$ {\mathcal R} = N(0.02\leq M/M_\odot\leq0.08)/N(0.08<M/M_\odot\leq10)$$

\begin{figure*}[t]
\begin{center}
\includegraphics[width=0.83\textwidth]{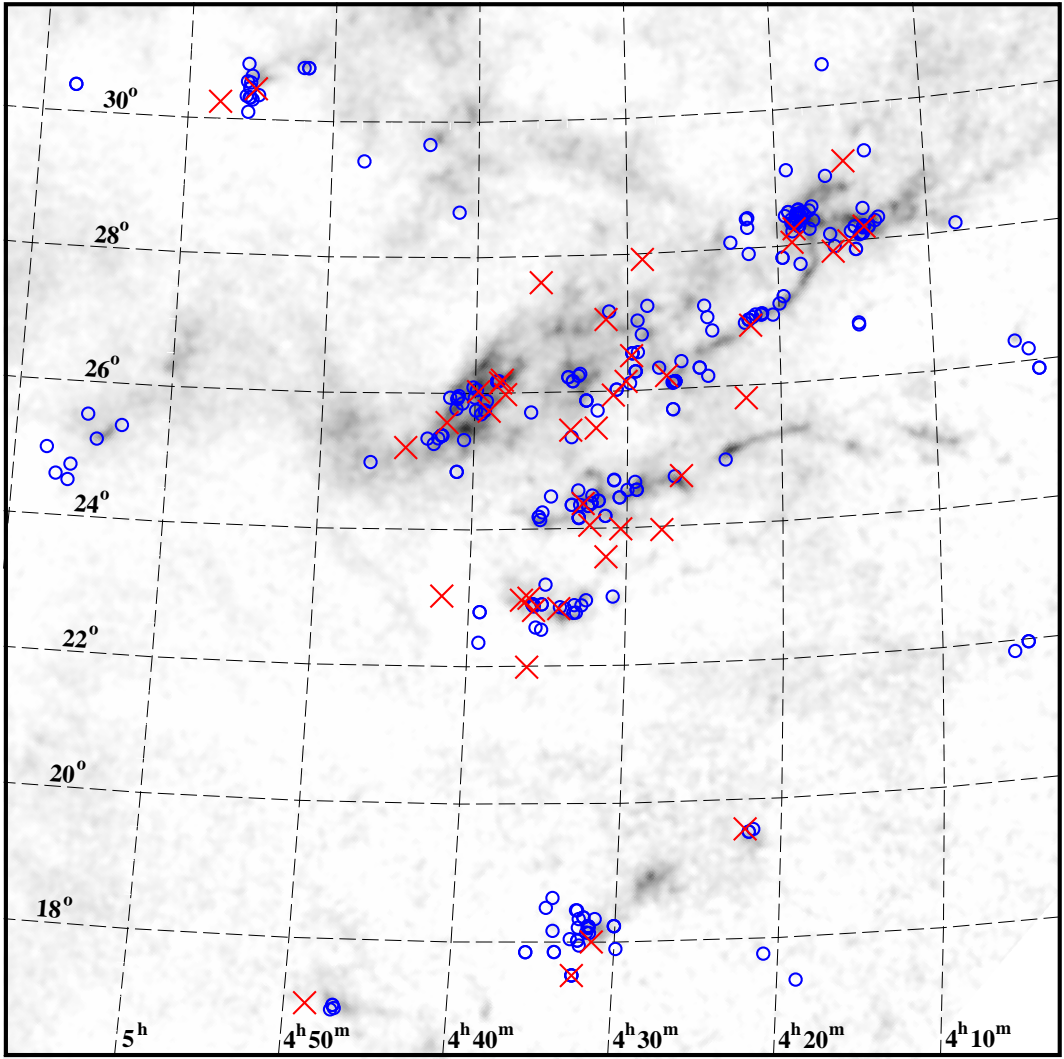}
\end{center}
\caption{\small 
Spatial distributions of stars ($\leq$M6, {\it circles}) and 
BDs ($>$M6, {\it crosses}) in the Taurus star-forming region 
shown with a map of extinction ({\it grayscale}, {\em Dobashi et al.}, 2005).
\label{fig:map}}
\end{figure*}

\noindent
The IMFs for Taurus, IC~348, and Orion exhibit 
${\mathcal R} = 0.18\pm0.04$, $0.12\pm0.03$, and $0.26\pm0.04$, respectively.
Because the IMF measurement for Chamaeleon~I is preliminary,
a reliable BD fraction is not yet available.
These BD fractions for Taurus and IC~348 
are a factor of two lower than the value for Orion.
However, upon spectroscopy of a large sample of
BD candidates in the Trapezium, {\em Slesnick et al.} (2004)
found a population of faint objects with stellar masses, possibly seen 
in scattered light, which had contaminated previous photometric IMF samples
and resulted in overestimates of the BD fraction in this cluster. 
After they corrected for this contamination,
the BD fraction in the Trapezium was a factor of only $\sim1.4$ higher
than the value in Taurus from {\em Luhman} (2004c). 
Through a survey of additional areas of Taurus, {\em Guieu et al.} (2006) 
have recently discovered 17 new low-mass stars and BDs. By combining these
data with the previous surveys, they measured a BD fraction that is still
higher, and thus closer to the value for Orion. However, 
{\em Luhman} (2006) finds that their higher BD fraction is
due to a systematic offset between the spectral types of 
{\em Guieu et al.} (2006) and the classification system used for the 
previously known late-type members of Taurus ({\em Luhman}, 1999; 
{\em Brice\~no et al.}, 2002). 
In summary, according to the best available data, the BD fractions
in Taurus and IC~348 are lower than in the Trapezium, but by a factor 
that is smaller than that reported in earlier studies.

Because the mass-luminosity relation is a function of age for BDs
at any age, and the ages of individual field BDs are unknown,
a unique, well-sampled IMF of field BDs cannot be constructed.
When substellar mass functions are instead compared in terms of
power-law slopes (Salpeter is 1.35), the latest constraints in the field from 
{\em Chabrier} (2002) ($\alpha\lesssim0$) and {\em Allen et al.} (2005) 
($-1.5\lesssim\alpha\lesssim0$) are consistent with the mildly
negative slopes exhibited by the data for star-forming regions
in Fig.~\ref{fig:imf}. Thus, data for both star-forming regions and 
the solar neighborhood are consistent with stars outnumbering BDs by a 
factor of $\sim5$-8.
If BDs form through ejection and have higher velocity dispersions than 
stars as predicted by {\em Kroupa and Bouvier} (2003) (but not 
{\em Bate et al.}, 2003), then the BD fraction would be higher in the 
field than in star-forming regions since BD members would be quickly ejected
from the latter. However, current data show no evidence of such a difference. 

In addition to the abundance of BDs relative to stars, the minimum mass
at which BDs can form also represents a fundamental constraint for theories
of BD formation. For several of the star-forming regions cited in this section, 
BDs with conclusive evidence of membership and accurate spectral classifications
have been discovered down to optical spectral types of M9.5,
corresponding to masses of $\sim10$-20~$M_{\rm Jup}$. 
Additional BDs have been reported at cooler and fainter levels, most notably
in $\sigma$~Ori. However, some of these objects lack clear evidence of
membership and instead could be field dwarfs
({\em Burgasser et al.}, 2004, references therein).
Finally, {\em Kirkpatrick et al.} (2006) recently discovered a young L dwarf in
the field that is probably comparable in mass (6-25~$M_{\rm Jup}$) to the least
massive BDs found in young clusters.

\section{\textbf{KINEMATICS AND POSITIONS AT BIRTH}}

Some models for the formation of BDs via embryo ejection predict that 
BDs are born with higher velocity dispersions than stars and thus are more
widely distributed in star-forming regions than their 
stellar counterparts ({\em Reipurth and Clarke}, 2001; 
{\em Kroupa and Bouvier}, 2003).
Meanwhile, other models of ejection ({\em Bate et al.}, 2003) 
and models in which BDs form in a star-like manner predict that 
stars and BDs should have similar spatial and velocity distributions.
Because normal dynamical evolution of a cluster can produce mass-dependent
distributions like those of the first set of ejection models
({\em Bonnell and Davies}, 1998), clusters that are old or dense are
not suitable for testing their predictions ({\em Moraux and Clarke}, 2005).
Therefore, based on their youth and low stellar densities, the Taurus 
and Chamaeleon star-forming region are ideal sites for comparing the 
positions and kinematics of stars and BDs.

Precise radial velocities of low-mass stars and BDs in Chamaeleon~I 
measured from high-resolution spectra are slightly
less dispersed (0.9$\pm$0.3\,km/s) but still consistent
with those of stars (1.3$\pm$0.3\,km/s) 
({\em Joergens and Guenther}, 2001; {\em Joergens}, 2006b).
The BDs do not show a
high velocity tail as predicted by some models of the ejection
scenario ({\em Sterzik and Durisen}, 2003; {\em Umbreit et al.}, 2005).
Similar results have been found for Taurus ({\em White and Basri}, 2003;
{\em Joergens}, 2006b).
While the absence of a significant mass dependence of
the velocities is consistent with some models of the ejection
scenario ({\em Bate et al.}, 2003; {\em Delgado-Donate et al.}, 2004),
the observed global radial velocity dispersion (BDs and stars)
for Chamaeleon~I members is smaller than predicted by any model
of the ejection scenario.

Over time, the surveys for BDs in Taurus have encompassed steadily larger 
areas surrounding the stellar aggregates (see references in previous section).
These data have exhibited no statistically significant differences in the
spatial distribution of the high- and low-mass members of Taurus 
({\em Brice\~no et al.}, 2002; {\em Luhman}, 2004c; {\em Guieu et al.}, 2006).
This result has been established definitively by the completion of a BD survey 
of 225~deg$^2$ encompassing all of Taurus ({\em Luhman}, 2006). 
As shown in Fig.~\ref{fig:map}, the BDs 
follow the spatial distribution of the stellar members, 
and there is no evidence of a large, distributed population of BDs.
As with the kinematic properties, these spatial data are consistent with a
common formation mechanism for stars 
and BDs and some models for ejection ({\em Bate et al.}, 2003), but
not others ({\em Kroupa and Bouvier}, 2003).

\section{\textbf{MULTIPLICITY}}

As with stars, the multiplicity properties of BDs (frequency, separation, 
and mass ratio distributions) are intimately tied to their formation. 
As discussed in the chapter by {\em Whitworth et al.}, embryo-ejection 
scenarios predict few binaries and only close orbits, while isolated
fragmentation models allow for higher binary frequencies and larger 
maximum separations. Therefore, accurately characterizing the multiplicity of
BDs can help distinguish between these scenarios (and others).
Moreover, the identification of very low-mass companions to BDs will delimit 
better the types of environments in which planets can form.
The chapter by {\em Burgasser et al.} provides a comprehensive review of
the observational and theoretical work on the binary properties of BDs.
In this section, we discuss highlights of the latest
observational work and their implications for the origin of BDs.

\subsection{\textbf{Brown Dwarf Companions to Stars: The Brown Dwarf Desert}}

Among companions at separations less than a few AU from solar-type stars, radial
velocity surveys have revealed a paucity of BDs (20-80~$M_{\rm Jup}$)
relative to giant planets and stellar companions ({\em Marcy and Butler}, 2000).
Deficiencies in substellar companions have been observed
at wider separations as well, as illustrated
in Fig.~8 from {\em McCarthy and Zuckerman} (2004),
which compared published frequencies of
stellar and substellar companions as a function of separation.
At separations less than 3~AU, the frequency of BD 
companions is $\sim0.1$\% ($<0.5$\%) 
({\em Marcy and Butler}, 2000) and the frequency of stellar companions 
is $13\pm3$\% ({\em Duquennoy and Mayor}, 1991; {\em Mazeh et al.}, 1992), 
indicating that BDs are outnumbered by stars among close companions by a factor
of $\sim100$ ($>20$).
In comparison, the ratio of the frequencies of stellar and 
substellar companions is between $\sim3$ and 10 at wider separations 
({\em McCarthy and Zuckerman}, 2004), which is comparable to the ratio of the 
numbers of stars and BDs in isolation ($\sim5$-8, Section~\ref{sec:imf}).
Thus, for solar-type primaries, only the close companions exhibit a true desert
of BDs. The similarity in the abundances of BDs among wider companions 
and free-floating objects suggests that they arise from 
a common formation mechanism (e.g., core fragmentation.)

\subsection{\textbf{Binary Brown Dwarfs}}

\begin{figure}[t]
\begin{center}
\includegraphics[width=.48\textwidth]{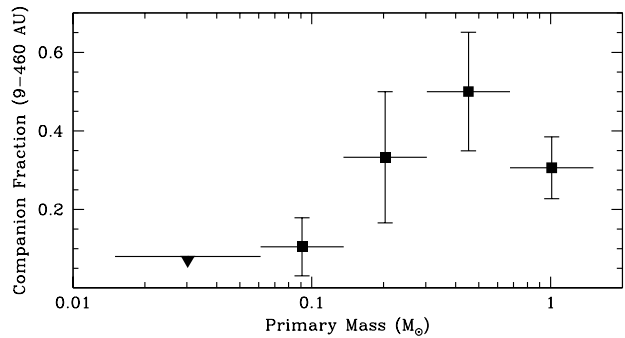}
\end{center}
\caption
{\label{fig2}
\small{Binary fraction over the
separation range 9-460 AU versus primary mass for young stars and
BDs in the Taurus star-forming region ({\em White et al.}, in preparation;
{\em Kraus et al.}, 2006).
}
}
\end{figure}

Binary surveys of members of the solar neighborhood 
have found progressively smaller binary fractions,
smaller average and maximum separations, and larger mass ratios 
($q\equiv M_2/M_1$) with decreasing primary mass from stars to BDs
({\em Duquennoy and Mayor}, 1991; {\em Fischer and Marcy}, 1992;
{\em Reid et al.}, 2001; {\em Bouy et al.}, 2003;
{\em Burgasser et al.}, 2003; {\em Close et al.}, 2003; {\em Gizis et al.},
2003; {\em Siegler et al.}, 2005).
To help identify the sources of these trends (e.g., formation mechanism, 
environment), it is useful to compare the field data to 
measurements in young clusters and star-forming regions.
Because young clusters have greater distances than the nearest stars and
BDs in the field, the range of separations probed in binary surveys of
young clusters is usually smaller than that for field objects. 
As a result, accurate measurements of the multiplicity as a function of
mass are difficult for young populations. However, for one of the best
studied star-forming regions, Taurus-Auriga, 
{\em White et al.} (in preparation) and {\em Kraus et al.} (2006) 
have measured the binary fraction (for $a=9$-460~AU, $q\ge0.09$, 
defined for completeness) as a function of mass from 
1.5 to 0.015~$M_\odot$. As shown in Fig.~\ref{fig2}, the binary fraction 
in Taurus declines steadily with primary mass, which resembles the
trend observed for the solar neighborhood.
This behavior can be explained by simple random pairing from the same 
mass function without the presence of different formation mechanisms 
at high and low masses.
Out of the 17 Taurus members with spectral types cooler than
M6 ($\lesssim0.1$~$M_\odot$), none have spatially resolved companions,
which again is consistent with the small separations of $a<20$~AU
that have been observed for most of the binary low-mass stars and BDs
in the field. A similar paucity of wide low-mass pairs has been found in other
young regions, including IC~348 ({\em Duch\^ene et al.}, 1999; 
{\em Luhman et al.}, 2005c), Chamaeleon~I ({\em Neuh\"{a}user et al.}, 2002),
Corona Australis ({\em Bouy et al.}, 2004), Upper Scorpius
({\em Kraus et al.}, 2005), and the Trapezium Cluster in
Orion ({\em Lucas et al.}, 2005).
However, in both the field and in young clusters, a few wide binary
low-mass stars and BDs have been found at projected separations 
that range from 33-41~AU ({\em Harrington et al.}, 1974; 
{\em Mart{\'\i}n et al.}, 2000; {\em Chauvin et al.}, 2004, 2005; 
{\em Phan-Bao et al.}, 2005) to beyond 100~AU ({\em White et al.}, 1999;
{\em Gizis et al.}, 2001; {\em Luhman}, 2004b, 2005; {\em Bill\`{e}res et al.}, 
2005; {\em Bouy et al.}, 2006). 
Because these wide binaries are weakly bound and extremely fragile,
it would seem unlikely that they have been subjected to violent dynamical 
interactions, suggesting that some low-mass stars and BDs are able
to form without the involvement of ejection, apparently through
standard, unperturbed cloud fragmentation.
Indeed, in embryo-ejection simulations, {\em Bate et al.} (2002) found that 
"because close dynamical interactions are involved in their formation...binary
brown dwarf systems that do exist must be close, $\lesssim10$~AU".
However, in more recent calculations by {\em Bate and Bonnell} (2005),
a wide binary BD was able to form when two BDs were simultaneously
ejected in similar directions.

\begin{figure}[t]
\begin{center}
\includegraphics[width=0.5\textwidth]{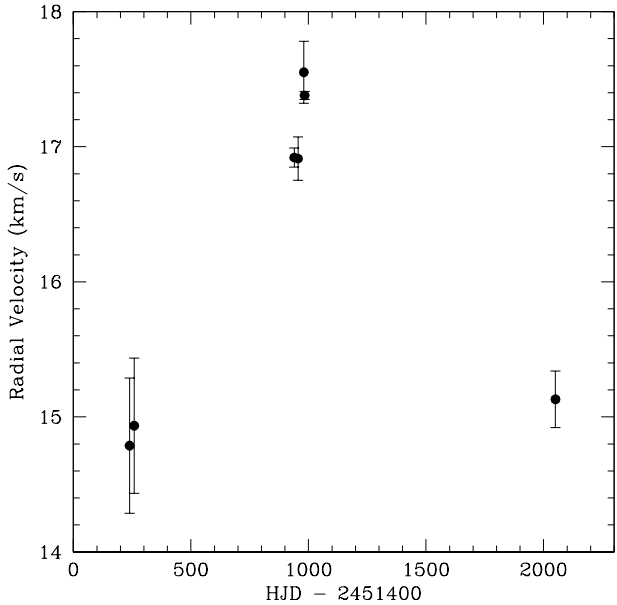}
\end{center}
\caption
{\label{cha8}
\small{Radial velocity data for the young low-mass object
Cha\,H$\alpha$\,8 (M6.5) recorded with UVES/VLT: significant variability 
occurring on time scales of months to years hint at a companion at $a>$0.2~AU
and M~$\sin i\gtrsim6$~$M_{\rm Jup}$ ({\em Joergens}, 2006a).
}
}
\end{figure}

Because the surveys for binary BDs cited above employed direct 
imaging, they were not sensitive to very close binaries ($a\lesssim1$ and 
$a\lesssim10$~AU for the field and clusters, respectively), 
making the resulting binary fractions only lower limits.
Spectroscopic monitoring for radial velocity variations provides a means of 
identifying the closest companions, which is essential for assessing whether
the formation mechanism of companions in substellar multiple systems 
changes with separation. The first free-floating BD to be discovered, 
PPL~15 ({\em Stauffer et al.}, 1994), turned out to be a spectroscopic 
binary with companions of nearly equal mass in a 6 day orbit 
({\em Basri and Mart\'{\i}n}, 1999). More recently,
{\em Guenther and Wuchterl} (2003) started a systematic survey for close 
companions to 25 low-mass stars and BDs in the field, finding 
two candidate double-lined spectroscopic binaries.
An additional object in their sample was found to be a binary by 
{\em Reid et al.} (2002).
{\em Joergens and Guenther} (2001) started a similar survey for close 
low-mass binaries in the Chamaeleon~I star-forming region. 
Among a subsample of ten low-mass objects ($M\la$0.12~$M_{\odot}$, M5--M8),
none show signs of companions down to the masses of giant planets
for orbital periods $P<40$~d, corresponding to separations of $a<0.1$~AU 
({\em Joergens}, 2006a). For Cha~H$\alpha$~8 (M6.5), data recorded across a 
longer period of time does indicate the existence of a spectroscopic companion 
of planetary or BD mass with an orbital period of several months to a few years,
as shown in Fig.~\ref{cha8}.  In a combination of the above old and 
young samples, 3 ($\sim9$\%) and 4 ($\sim11$\%) objects have possible 
companions at $P<100$~d and $P<$1000~d, respectively.
For comparison, the frequencies of binaries among solar-type field stars
are 7\% and 13\% in these same period ranges ({\em Duquennoy and Mayor}, 1991).

\section{\textbf{ACCRETION}}

The past 10-15 years has seen the establishment of a disk accretion
paradigm in low-mass T Tauri stars that explains many of their
observed characteristics. The picture centers on the concept
of magnetospheric accretion, whereby the stellar magnetic field
truncates the circumstellar disk and channels accreting material
out of the disk plane and onto the star (see the chapter by 
{\em Bouvier et al.} and references therein).  
Models of magnetospheric accretion
successfully describe many features of classical T Tauri stars (CTTSs)
including the broad, asymmetric permitted line emission (e.g., {\em Muzerolle
et al.}, 2001) and blue/UV continuum excess 
(e.g., {\em Calvet and Gullbring}, 1998). 
The investigation of accretion and disk signatures in lower-mass objects
extending below the substellar limit is a natural extension of
the CTTS studies, helping to address the origin of BDs.

Accretion in young BDs is fundamental to
our understanding of formation mechanisms, and hence has undergone
considerable scrutiny following the discovery of the first substellar objects
in nearby star-forming regions. Spectroscopy of the first known
young BDs (e.g., {\em Luhman et al.}, 1997; 
{\em Brice\~no et al.}, 1998; {\em Com\'eron et al.}, 1999) 
revealed that many were superficially similar
to CTTSs in terms of emission line activity.  In particular, equivalent widths 
of H$\alpha$ emission in many cases exceeded levels typical of chromospheric
activity in low-mass CTTSs and main sequence dMe stars, suggesting the presence
of accretion. With the advent of 8-10~m class telescopes, high-resolution 
optical spectroscopy of young BDs became possible.  As a result, 
research by many groups over the last 5 years has provided conclusive 
evidence of ongoing accretion in many young substellar systems.  
This evidence includes
the presence of broad, asymmetric Balmer line profiles, continuum veiling
of photospheric absorption features, and in a few cases forbidden line
emission, all similar to features seen in CTTSs.

\subsection{\textbf{Diagnostics}}

The first demonstration of accretion infall in a very low-mass object
was presented by {\em Muzerolle et al.} (2000) for the Taurus member 
V410~Anon~13 ($\sim0.1$~$M_\odot$, {\em Brice\~no et al.}, 2002).
The H$\alpha$ profile for this object shows a clear
infall asymmetry similar to that commonly seen in CTTSs, albeit with
a narrower line width and a lack of opacity-broadened wings.  Such features
indicated ballistic infall at velocities consistent with the object's
mass and radius, and a much lower mass accretion rate ($\mdot$)
than typical of higher-mass CTTSs with similar ages.
Modeling of the profile in fact yielded an extremely small value of
$\mdot \sim 5 \times 10^{-12} \; \msunyr$,
a mere trickle in comparison with the average rate of $\sim10^{-8} \; \msunyr$
for solar-mass CTTSs ({\em Gullbring et al.}, 1998).

\begin{figure}[t]
\begin{center}
\includegraphics[width=.5\textwidth]{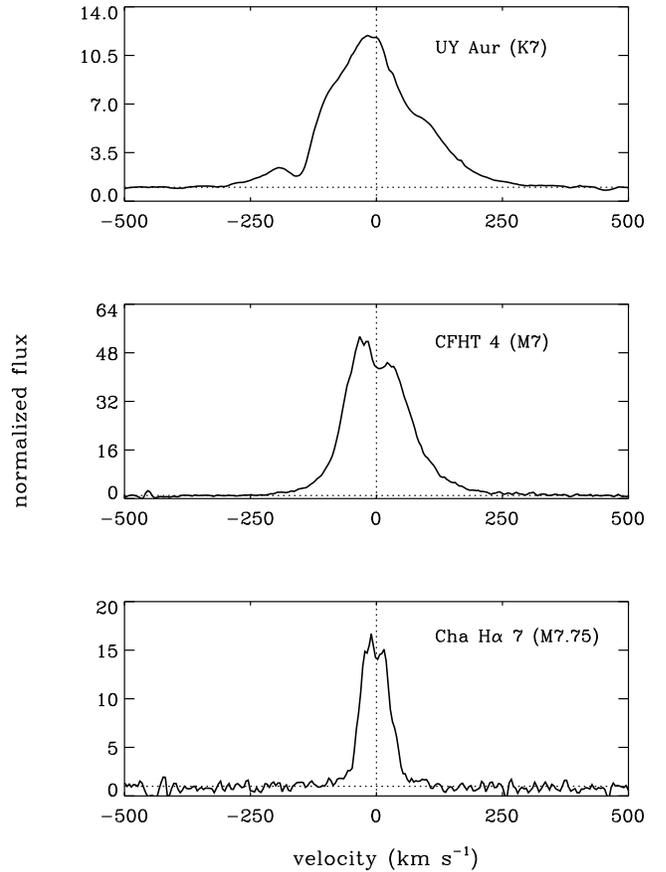}
\end{center}
\caption{\small Comparison of H$\alpha$ profiles of (from top to bottom)
a typical CTTS ({\em Muzerolle et al.}, 1998), a typical substellar accretor
(note the smaller line width indicative of the much smaller gravitational
potential), and a substellar non-accretor exhibiting the narrow and
symmetric profile produced by chromospheric emission
({\em Muzerolle et al.}, 2005).
\label{profiles}}
\end{figure}

Evidence for accretion in many other very low-mass stars and BDs
has since accumulated by various techniques. {\em White and Basri} (2003)
were the first to publish measurements of continuum veiling from
accretion shock emission from objects near and below the substellar limit,
providing more direct measures of mass accretion rates that were again
lower than the typical of CTTSs.  However, measurable veiling has turned out to 
be very rare in substellar accretors because of their small accretion rates.
Models of substellar accretion shock emission
({\em Muzerolle et al.}, 2000) show that measurable veiling is produced only
when $\mdot > 10^{-10} \; \msunyr$. 
Since H$\alpha$ emission from the accretion flow is detectable at much lower
$\mdot$, H$\alpha$ emission line profiles remain
the most sensitive accretion diagnostics available for young BDs.
Chromospheric emission associated with magnetic activity appears to be
a common feature of both young and older field dwarfs
(e.g., {\em Mohanty and Basri}, 2003), producing generally larger H$\alpha$
equivalent widths at later spectral types as a result of decreasing contrast
with the photosphere; H$\alpha$ equivalent widths of 20~\AA\ are not
uncommon in young objects with spectral types M5 or later.
Spectral type-dependent equivalent width criteria have been proposed
by several authors ({\em Barrado y Navascu\'es and Mart\'in}, 2003;
{\em White and Basri}, 2003).  However, line profiles offer
the most unambiguous discriminant, as chromospheric emission produces
much narrower and symmetric profiles compared to the broader and often
asymmetric accretion profiles, as illustrated in Fig.~\ref{profiles}.

A popular high-resolution accretion criterion has been the H$\alpha$
10\% line velocity width.  Accretion profile widths are related to
the maximum ballistic infall velocity
$V_{inf} \sim \sqrt{{2GM_* \over R_*} (1-{1 \over R_m})} \sim 160$ km s$^{-1}$
for $M_*=0.05 \; \msun$, $R_*=0.5 \; R_{\odot}$,
and a magnetospheric truncation radius $R_m = 3 \; R_*$
(see {\em Muzerolle et al.}, 2003). 
Most chromospheric profiles generally exhibit
velocity half-widths $\lesssim 70$ km s$^{-1}$, much lower than
the characteristic infall velocity.  Broadening from rapid rotation can
create larger line widths, as has been observed, but this is rare
(typical rotation velocities are $v\sin i \lesssim 20$ km s$^{-1}$:
{\em Muzerolle et al.}, 2003, 2005) and in any case can be checked
with v~sin~i measurements.  Also, objects with
large H$\alpha$ velocity widths consistent with infall tend to have
larger H$\alpha$ {\it equivalent} widths than those with chromospheric
profiles and also tend to correlate with the presence of other accretion
signatures such as other permitted and forbidden emission lines.
Adopting a 10\% line width threshold of $V_{10} \gtrsim 180-200$ km s$^{-1}$
gives reasonably accurate accretor identifications in BDs
({\em Jayawardhana et al.}, 2003b; {\em Muzerolle et al.}, 2005), 
although occasional misidentifications can occur for rapidly rotating 
nonaccretors or pole-on accretors.

Dozens of substellar accretors have now been identified down to masses 
approaching the deuterium burning limit and with ages from 1 to 10 Myr
(e.g., {\em Jayawardhana et al.}, 2003b; {\em Mohanty et al.}, 2005;
{\em Muzerolle et al.}, 2003, 2005).  Such a statistically robust sample has
allowed systematic studies of accretion properties across nearly the entire
range of substellar masses yet identified.
For instance, magnetospheric accretion requires
disk material to be present at or within the corotation radius,
which should be detectable at near- or mid-IR wavelengths.
Comparing known substellar accretors in Chamaeleon~I and Ophiuchus 
with IR excesses detected by ISO at 6.7 and 
14.3~\micron\ ({\em Natta et al.}, 2004) and Spitzer at
3.6-8~\micron\ ({\em Luhman et al.}, 2005d),
3/10 and 7/10 in each region, respectively,
exhibit both accretion and disks, while 3/10 in each region show disks
but no accretion.  There are no cases of accretion without disk signatures.
The objects with disks and lacking accretion signatures may simply
be accreting at rates below the observable threshold.
The larger fraction of these in Chamaeleon~I compared to Ophiuchus may be
a reflection of the slightly older age of the former, so that the disks
have evolved to lower accretion rates on average (see below).
Indeed, many studies have now
found strong indications of a decreasing fraction of accreting objects
with age, both above and below the substellar limit.
Typical values range from 30-60\% in 1-3 Myr-old regions such as
Taurus and Chamaeleon~I, but drop significantly to 0-5\%
in 3-5 Myr-old regions such as $\sigma$~Ori and Upper Scorpius
({\em Muzerolle et al.}, 2005; {\em Mohanty et al.}, 2005).
These numbers are consistent with similar declines observed in
the accretor fraction at stellar masses, indicating similar evolutionary 
timescales for accretion between stars and BDs.
The same result is found for the disk fractions of stars and BDs,
as we discuss in Section~\ref{sec:diskfractions}.

\subsection{\textbf{Substellar Accretion Rates}}
\label{sec:rates}

\begin{figure}[h]
\begin{center}
\includegraphics[width=.5\textwidth]{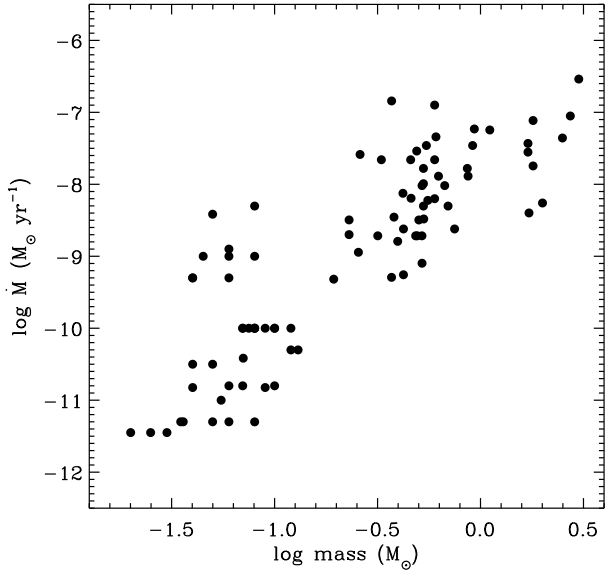}
\end{center}
\caption{\small Mass accretion rate as a function of substellar and stellar
mass for objects in Taurus (1~Myr), Cha~I (2~Myr), IC~348 (2~Myr), and 
Ophiuchus (0.5~Myr) ({\em Gullbring et al.}, 1998; {\em White and Ghez}, 2001;
{\em Muzerolle et al.}, 2000, 2003, 2005; {\em Natta et al.}, 2004;
{\em Mohanty et al.}, 2005). These regions exhibit similar accretion rates
at a given mass, except for slightly higher rates in Ophiuchus.
\label{mass_mdot}}
\end{figure}

The results summarized above show that the overall accretion characteristics
are essentially continuous across the substellar boundary, which is 
consistent with stars and BDs forming via the same accretion processes.  
A more quantitative assessment can be made from measurements of mass accretion 
rates. Most of the estimates of $\mdot$ based on line profile
modeling ({\em Muzerolle et al.}, 2003, 2005) and secondary IR calibrators
such as Paschen $\beta$ and Brackett $\gamma$ ({\em Natta et al.}, 2004)
and the Ca II triplet ({\em Mohanty et al.}, 2005) have shown that very small
accretion rates are in fact typical of very low-mass young objects.
The average value for substellar accretors is roughly 2-3 orders of
magnitude lower than that of 1 Myr-old CTTSs.  A clear trend
of decreasing accretion rate with decreasing mass was found by
{\em Muzerolle et al.} (2003) and subsequently extended down to 
$M\sim0.02$~$\msun$ by {\em Mohanty et al.} (2005) and {\em Muzerolle et al.} 
(2005), with a functional form of $\mdot \propto M^2$ (Fig.~\ref{mass_mdot}).
The surprising correlation between mass and accretion rate has profound
implications for BD origins.
The lack of any obvious shift in the correlation about the substellar limit
implies a continuity in the formation processes of stars and BDs.
This lends support to BD formation via fragmentation and collapse
of low-mass cloud cores.  
However, the physical origins of accretion in young stellar objects
need to be better understood before definitive conclusions can be made.

What is the source of the mass-accretion correlation?
The general theory of viscously accreting disks
does not predict a strong relation between these two quantities.
The answer may lie in the essentially unknown source of viscosity
needed to drive accretion.  A commonly invoked mechanism is
the Balbus-Hawley instability, which requires sufficient ionization of
disk gas to effectively couple with magnetic fields.  X-ray activity
from the stellar magnetic field is a potential ionization source
({\em Glassgold et al.}, 2004); {\em Muzerolle et al.} (2003) suggested 
that the observed correlation $L_X \propto M_*^2$ 
(e.g., {\em Feigelson et al.}, 2003) may then be related to
the similar dependence of accretion rate on mass.
However, a comparison between $L_X$ and $\mdot$ for the small number
of objects for which both quantities have been measured reveals no
statistically significant correlation, although the mass range
covered is not very large.  More observations of both quantities
are needed; a particularly interesting analysis would be to compare
accretion variability at the onset of and subsequent to an X-ray flare event.

More recently, {\em Padoan et al.} (2005) have proposed a modified
Bondi-Hoyle accretion model in which young stars and BDs
accrete primarily from the large-scale medium in which they are moving
rather than from their disks alone. In this case, the accretion rate can be
determined by the density and sound speed of the surrounding gas
and the relative velocity between the object and that material.
The resultant relation produces the correct mass dependence.
However, it is not clear how a Bondi-Hoyle flow would interact with the disk.
For instance, it may be incorporated into the disk prior to reaching the star;
note that the observed accretion diagnostics are inconsistent with
spherical infall onto the stellar surface.
If so, the Bondi-Hoyle relation may in fact
determine the rate of residual infall onto the disk, but not necessarily
the rate of accretion onto the star, which is what is measured.
In addition, the model cannot explain accretors located in low-density
regions far from molecular clouds, such as the well-known CTTS TW Hydrae.
Comparisons of mass accretion rates versus surrounding cloud 
temperatures and densities need to be made to further assess
the applicability of this model.

\subsection{\textbf{Jets and Outflows}}

Many other similarities in accretion activity between stars and BDs
have been found, including photometric and line profile
variability ({\em Caballero et al.}, 2004; 
{\em Scholz and Eisloffel}, 2004, 2005; {\em Scholz et al.}, 2005), detections 
of H$_2$ emission in the UV ({\em Gizis et al.}, 2005), possible detections of
UV continuum excesses ({\em McGehee et al.}, 2005), and evidence of 
accretion-generated outflows such as blueshifted absorption and forbidden 
emission ({\em Fern\'andez and Comer\'on}, 2001; {\em Muzerolle et al.}, 2003; 
{\em Barrado y Navascu\'es et al.}, 2004a; {\em Luhman}, 2004c;
{\em Mohanty et al.}, 2005).
Among the four accretors at M6 or later from {\em Muzerolle et al.} (2003), 
one shows forbidden line emission, while the two Class~I objects at M6 
from {\em White and Hillenbrand} (2004) show forbidden line emission.  
Based on small number statistics, jet signatures appear to be
less often associated with accretion signatures for low-mass stars and 
BDs than for stars. However, this may stem from 
lower mass loss rates in substellar jets producing emission that is 
more difficult to detect ({\em Masciadri and Raga}, 2004).
{\em White and Hillenbrand} (2004) found that the ratio of mass loss to mass
accretion rate is the same for objects with both high and low mass accretion 
rates, though with considerable dispersion.
Thus, the low accretion rates inferred for BDs 
likely correspond to diminished mass loss rates and less luminous
forbidden line emission, possibly below typical detection levels.  
Overall, the sparse data on jets from young
accreting BDs are similar to those of higher mass CTTSs,
but on a smaller and less energetic scale. 

In addition to the above indirect evidence for outflows provided by 
forbidden line emission, {\em Whelan et al.} (2005) and {\em Bourke et al.}
(2005) have spatially resolved outflows toward ISO~102 in Ophiuchus
(also known as GY~202)
and L1014-IRS through optical forbidden lines and millimeter CO emission, 
respectively.  Although {\em Whelan et al.} (2005) referred to ISO~102
as a BD, the combination of its M6 spectral type from 
{\em Natta et al.} (2002), the evolutionary models of {\em Chabrier et al.}
(2000), and the temperature scale of {\em Luhman et al.} (2003b) suggest 
that it could have a stellar mass of $\sim0.1$~$M_\odot$. 
It appears likely that L1014-IRS has a substellar mass 
({\em Young et al.}, 2004; {\em Huard et al.}, 2006), but this is difficult to
confirm because of its highly embedded nature. The molecular outflow detected
toward this object by {\em Bourke et al.} (2005) is one of the smallest known
outflows in terms of its size, mass, and energetics.

\section{\textbf{CIRCUMSTELLAR DISKS}}

The collapse of a cloud core naturally produces a circumstellar disk via 
angular momentum conservation.
Thus, understanding the formation of BDs
requires close scrutiny of their circumstellar disks.
In addition, as with stars, studying disks around BDs should
provide insight into if and how planet formation occurs around these
small bodies. In this section, we summarize our knowledge of disks around  
BDs and discuss the resulting implications for the origin of BDs, 
the evolution of their disks, and the formation of planets.

\subsection{\textbf{Detections of Disks}}

Although resolved images of disks around BDs are not yet available, 
there is mounting evidence for their existence through detections of 
IR emission above that expected from stellar photospheres alone.
Modeling of the IR spectral energy distributions (SEDs) of young
BDs showing excess emission strongly suggests that the emitting
dust resides in disk configurations. For instance, 
{\em Pascucci et al.} (2003) considered different shell and disk geometries 
for a BD system in Taurus and demonstrated 
that spherically distributed dust with a mass estimated from the millimeter
measurements ({\em Klein et al.}, 2003) would produce much more extinction than
observed toward the BD. In comparison,
when the same material is modeled as a disk, the SED 
can be well reproduced without conflicting with the observed low extinction.

Excess emission in the $K$ and $L$ bands has been
observed for several young objects at M6-M8 and for a few as late as M8.5
({\em Luhman}, 1999, 2004c; {\em Lada et al.}, 2000, 2004; 
{\em Muench et al.}, 2001; {\em Liu et al.}, 2003; {\em Jayawardhana et 
al.}, 2003a). Excesses at longer, mid-IR wavelengths have been detected 
for CFHT~4 (M7, {\em Pascucci et al.}, 2003; {\em Apai et al.}, 2004), 
Cha~H$\alpha$~1 (M7.75, {\em Persi et al.}, 2000; {\em Comer\'on et al.}, 2000;
{\em Natta and Testi}, 2001; {\em Sterzik et al.}, 2004), 
2MASS~1207-3932 (M8, {\em Sterzik et al.}, 2004), 
GY141 (M8.5, {\em Comer\'on et al.}, 1998), and several late-type objects
in Ophiuchus ({\em Testi et al.}, 2002; {\em Natta et al.}, 2002;
{\em Mohanty et al.}, 2004).
{\em Klein et al.} (2003) has extended these detections of circumstellar 
material to millimeter wavelengths for CFHT~4 and IC~348-613 (M8.25).

\begin{figure}[h]
\begin{center}
\includegraphics[width=.5\textwidth]{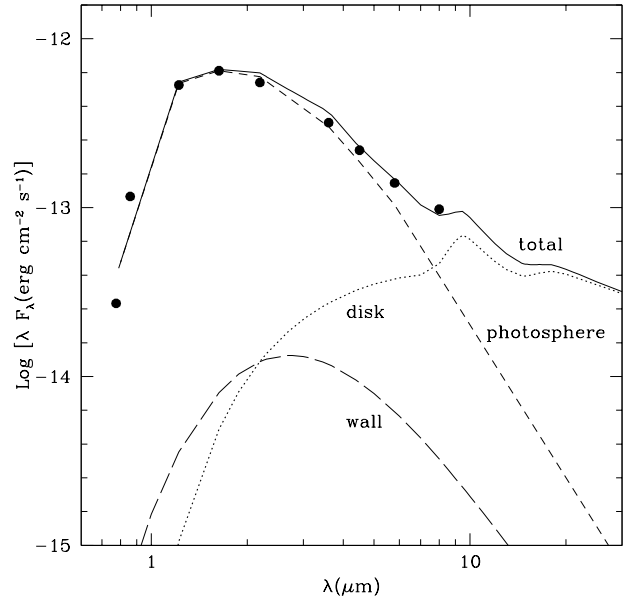}
\end{center}
\caption{\label{fig:cha208428}
\small 
SED of the least massive BD known to
harbor a disk, Cha~1109-7734 ({\it points}, {\em Luhman et al.}, 2005e). 
Relative to the distribution
expected for its photosphere ($\sim$M9.5, {\it short dashed line}), this BD
exhibits significant excess emission at wavelengths longer than
5~\micron. The excess flux is modeled in terms of emission from a circumstellar
accretion disk ({\it dotted line}) and a
vertical wall at the inner disk edge ({\it long dashed line}).
The sum of this disk model and the photosphere ({\it solid line}) is 
a reasonable match to the data for Cha~1109-7734.
}
\end{figure}

Because the Spitzer Space Telescope ({\em Werner et al.}, 2004) is far more 
sensitive beyond 3~\micron\ than any other existing facility, 
it is capable of detecting disks for BDs at very low masses.
To search for circumstellar disks around BDs at the lowest possible
masses, {\em Luhman et al.} (2005b) 
obtained mid-IR images (3.6-8~\micron) of the Chamaeleon~I star-forming region 
with the Infrared Array Camera (IRAC, {\em Fazio et al.}, 2004) on Spitzer.
In these data, they detected mid-IR excess emission from 
the coolest and least massive known BD in the cluster,
OTS~44 ({\em Oasa et al.}, 1999), which has a spectral type of $\gtrsim$M9.5 
and a mass of $M\sim15$~$M_{\rm Jup}$ ({\em Luhman et al.}, 2004).
By obtaining even deeper IRAC images of Chamaeleon~I and combining them
with optical and near-IR images from the Hubble Space Telescope and 
the CTIO 4~m, {\em Luhman et al.} (2005e) discovered a BD that is twice as 
faint as OTS~44 and exhibits mid-IR excess emission. 
By comparing the bolometric luminosity of this object, Cha~1109-7734, to
the luminosities predicted by the evolutionary models of {\em Chabrier et al.}
(2000) and {\em Burrows et al.} (1997), {\em Luhman et al.} (2005e)
estimated a mass of $8^{+7}_{-3}$~$M_{\rm Jup}$, placing it within the mass
range observed for extrasolar planetary companions 
($M\lesssim15$~$M_{\rm Jup}$, {\em Marcy et al.}, 2005).
{\em Luhman et al.} (2005e) successfully modeled the mid-IR excess emission
for Cha~1109-7734 in terms of an irradiated viscous accretion disk with
$\dot{M}\lesssim10^{-12}$~$\rm M_{\odot}\,yr^{-1}$, as shown in 
Fig.~\ref{fig:cha208428}, making it the least massive BD observed to have
a circumstellar disk and demonstrating that the basic ingredients for making 
planets are present around free-floating planetary-mass bodies.

\subsection{\textbf{Disk Fractions and Lifetimes}}
\label{sec:diskfractions}

Extensive work has been done in measuring disk fractions for stars
(e.g., {\em Kenyon and Hartmann}, 1995; {\em Hillenbrand et al.}, 1998;
{\em Haisch et al.}, 2001), which typically consists of IR photometry of a 
significant fraction of a young stellar population and identification of the
objects with excess emission.  Attempts have been made to extend measurements 
of this kind to low-mass stars and BDs.
Using $JHKL\arcmin$ photometry, {\em Jayawardhana et al.} (2003a) 
searched for excess emission among 53 objects in IC~348, Taurus, 
$\sigma$~Ori, Chamaeleon~I, the TW Hya association, Upper Scorpius, 
and Ophiuchus, 27 of which are later than M6 and thus likely to be substellar.
For the individual populations, the disk fractions for the stars and 
BDs exhibited large statistical errors of $\sim25$\%. 
For a sample combining Chamaeleon~I, IC~348, Taurus, and U~Sco, 
the number statistics were better, and {\em Jayawardhana et al.} (2003a) 
found a disk fraction of 40-60\%. 
Their disk/no disk classifications agreed well with those based on the 
Spitzer data from {\em Luhman et al.} (2005d) for types of $\leq$M6.
However, the two objects later than M6 in IC~348 and Chamaeleon~I 
that were reported to have disks by {\em Jayawardhana et al.} (2003a) 
show no excess emission in the Spitzer colors.  
{\em Liu et al.} (2003) also performed an $L\arcmin$-band survey
of low-mass objects. They considered a sample of 7 and 32 late-type members of
Taurus and IC~348, respectively, 12 of which have optical spectral types
later than M6. For their entire sample of low-mass stars and BDs,
{\em Liu et al.} (2003) found a disk fraction of $77\pm15$\%, which is a 
factor of two larger than measurements for IC~348 from Spitzer
({\em Luhman et al.}, 2005d). 9/10 objects with $E(K-L\arcmin)>0.2$
in the data from {\em Liu et al.} (2003) did exhibit significant excesses 
in the Spitzer colors, but the putative detections of disks with smaller 
$L\arcmin$ excesses were not confirmed by Spitzer. 
Any bona fide detection of a disk at $L\arcmin$ would be easily verified
with Spitzer given that the contrast of a disk relative to the central object
increases with longer wavelengths.

Because disks around BDs produce little $L\arcmin$-band emission compared to 
stellar systems, the $L\arcmin$-band surveys were not able to reliably detect 
BD disks. Meanwhile, BD disk excesses are larger at longer wavelengths, but 
measurements of this kind are feasible for only a small number of the 
brighter, more massive objects with most telescopes. 
In comparison, because Spitzer is highly sensitive and can survey large areas
of sky, it can reliably and efficiently detect disks for BDs at very
low masses and for large numbers of BDs in young clusters.
{\em Luhman et al.} (2005d) used IRAC on Spitzer to obtain mid-IR images of 
IC~348 and Chamaeleon~I, which encompassed 25 and 18 spectroscopically 
confirmed low-mass members
of the clusters, respectively ($>$M6, $M\lesssim0.08$~$M_\odot$).
They found that $42\pm13$\% and $50\pm17$\% of the two samples exhibit
excess emission indicative of circumstellar disks.
In comparison, the disk fractions for stellar members of these clusters
are $33\pm4$\% and $45\pm7$\%
(M0-M6, 0.7~$M_\odot\gtrsim M\gtrsim0.1$~$M_\odot$).
The similarity of the disk fractions of stars and BDs
indicates that the raw materials for planet formation are available
around BDs as often as around stars and 
supports the notion that stars and BDs share a common formation history.
However, as with the continuity of accretion rates from stars to BDs 
from Section~\ref{sec:rates}, these results do not completely exclude some 
scenarios in which BDs form through a distinct mechanism. For instance,
during formation through embryo ejection, the inner regions of disks 
that emit at mid-IR wavelengths could survive, although one might expect
these truncated disks to have shorter lifetimes than those around stars. 

When disk fractions for stellar populations across a range of ages 
(0.5-30~Myr) are compared, they indicate that the inner disks around stars 
have lifetimes of $\sim6$~Myr ({\em Haisch et al.}, 2001). 
Accurate measurements of disk fractions for BDs are available only for
IC~348 and Chamaeleon~I, both of which have ages near 2~Myr, and so
a comparable estimate of the disk lifetime for BDs is not currently possible.
However, the presence of a disk around a BD in the TW~Hya association
({\em Mohanty et al.}, 2003; {\em Sterzik et al.}, 2004), which has an age
of 10~Myr, does suggest that the lifetime of BD disks might be similar 
to that of stars. 
 
\subsection{\textbf{Disk Mass}}

A few estimates of dust masses of disks around BDs have been obtained through 
deep single-dish millimeter observations.
In a survey of 9 young BDs and 10 field BDs, 
{\em Klein et al.} (2003) detected disks around two of the young objects.
Because the millimeter emission is optically thin, fluxes could be 
converted to total disk masses assuming dust emission coefficients typical
to disks for low-mass stars and the standard gas-to-dust mass ratio of 100.
They derived disk masses of 0.4-6\,$M_{\rm Jup}$, which are a few percent of 
the BD masses, thus suggesting that disk masses scale with the 
mass of the central object down to the substellar regime. 
Similar measurements for larger samples of low-mass stars and BDs are 
needed to confirm such a trend.

\subsection{\textbf{Disk Geometry}}
\label{sec:geo}

Various theoretical studies have shown that the disk geometry strongly impacts 
the SED and have investigated the link between disk geometry and dust evolution 
(e.g., {\em Dullemond and Dominik}, 2005).
Flared disks are those with opening angles increasing with the disk 
radius as a consequence of vertical hydrostatic equilibrium 
({\em Kenyon and Hartmann}, 1987).
Such geometry characterizes the early phases of the disk 
evolution prior to dust processing (grain growth and dust settling, see
also the chapter by {\em Natta et al.}). 
Flatter disk geometries supposedly represent the evolutionary stage after 
flared disks (see also the chapter by {\em Dullemond et al.}).
Because flared disks intercept more stellar radiation than 
flat ones,  especially at large distances from the star, 
flared disks produce larger mid- and far-IR fluxes 
and a more prominent silicate emission feature than flat ones 
(e.g., {\em Chiang and Goldreich}, 1997).
{\em Walker et al.} (2004) calculated that, under the assumption of
vertical hydrostatic equilibrium, BD disks should be highly 
flared with disk scale heights three times larger than those derived 
for disks around CTTSs.
 
The work by {\em Natta and Testi} (2001) represents the first attempt 
to investigate the geometry of disks around low-mass objects.
The authors used scaled-down T Tauri disks to reproduce 
ISO mid-IR measurements of two low-mass stars 
(Cha~H$\alpha$~2 and 9, M5.25 and M5.5) and one BD
(Cha~H$\alpha$~1, M7.75) in the Chamaeleon~I star-forming region.
They considered passive flared and flat disks and made a number of simplifying 
assumptions following the method of {\em Chiang and Goldreich} (1997, 1999).
Passive disks are appropriate for BDs because of their very low accretion 
rates (Section~\ref{sec:rates}).
They concluded that models of flared disks were required to fit the SEDs
of these three objects.
A similar approach was used by the authors to
investigate a larger sample of nine low-mass stars and BDs in the Ophiuchus
star-forming region ({\em Testi et al.}, 2002; {\em Natta et al.}, 2002). 
A more careful inspection of the flared and flat disk predictions revealed
that two ISOCAM broad-band measurements were not always sufficient to 
determine the disk geometry.
{\em Apai et al.} (2002) used ground-based mid-IR narrowband
photometry to probe the silicate emission feature in the disk of 
Cha~H$\alpha$~2 and thus add an important new constraint to the disk models
from {\em Natta and Testi} (2001).
Their measurements ruled out the presence of strong silicate emission 
that was predicted by {\em Natta and Testi} (2001) and found that a 
flatter disk structure was required to fit the observed SED.

\begin{figure}[t]
\begin{center}
\includegraphics[width=.51\textwidth]{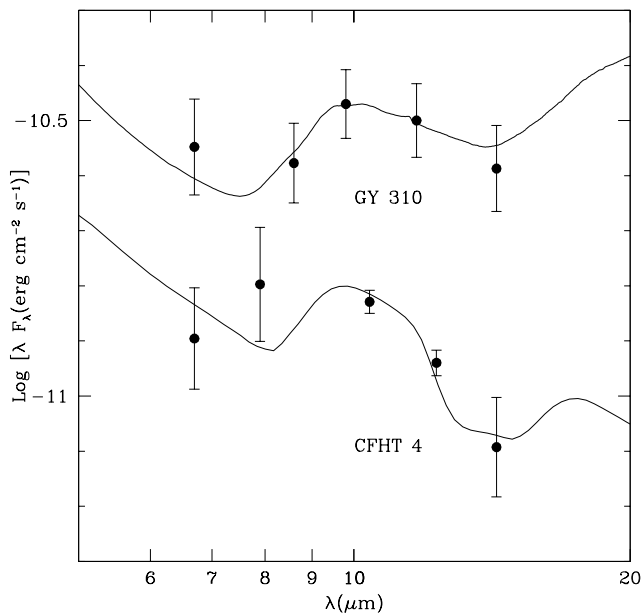}
\end{center}
\caption{\label{fig:sedc}
\small 
Comparison of geometries for two BD disks.
The best model fit for GY~310 consists of a flared disk with dust dominated by  
small sub-micron grains ({\em Mohanty et al.}, 2004). 
A good match to the SED of CFHT~4 can be achieved with a disk model 
with a little flaring and micron-sized grains ({\em Apai et al.}, 2004).
The SED of GY~310 has been shifted up by 0.5 dex.}
\end{figure}

Recent ground- and space-based measurements have provided more
comprehensive SEDs for about a dozen low-mass stars and BDs
({\em Pascucci et al.}, 2003; {\em Mohanty et al.}, 2004;
{\em Apai et al.}, 2004; {\em Sterzik et al.}, 2004; {\em Furlan et al.}, 2005a;
{\em Hartmann et al.}, 2005; {\em Muzerolle et al.}, 2006; 
{\em Pascucci et al.}, in preparation).
The modeling of these disks shows that flared, flat and intermediate flaring 
geometries all occur in BD disks (see Fig.~\ref{fig:sedc}).
A similar trend is found for disks around more massive stars
(e.g., {\em Furlan et al.}, 2005b). As with studies of disks at stellar masses, 
samples of BD disks from a greater variety of ages
and star-forming conditions are needed to distinguish between the
effects of evolution and environment on disk structure.

\subsection{\textbf{Dust Processing}}

Grain growth and dust settling are thought to represent the first steps of 
planet formation (e.g., {\em Henning et al.}, 2006). Studies of 
disks around intermediate-mass stars
also indicate a possible link between grain growth
and crystallinity, with high crystallinity measured in disks having grains 
larger than the dominant sub-micron interstellar grains 
(e.g., {\em van Boekel et al.}, 2005). 
Determining whether BD disks evolve into planetary systems
requires first identifying the presence of such dust processing.
Because dust settling is related to the disk geometry, some evidence of
dust processing can be gained by the kind of SED modeling described in 
the previous section. For instance, {\em Mohanty et al.} (2004)
concluded that the SED of a young BD in Ophiuchus, GY~310, was consistent
with a flared disk geometry and small interstellar grains, while
{\em Apai et al.} (2004) found that the SED of 
CFHT~4 was indicative of a flat disk structure 
(Fig.~\ref{fig:sedc}). The latter authors also found that 
the peak position of the silicate emission feature and the 
line-to-continuum flux ratio demonstrated that the emission was dominated by 
grains about 10 times larger than the dominant 0.1~\micron\ interstellar grains.
Fitting the emission and the overall continuum required a disk with 
intermediate flaring. This work indicated that young BD disks process dust 
in a similar fashion as disks around stars (e.g., {\em Przygodda et al.}, 2003).

\begin{figure}[t]
\begin{center}
\includegraphics[width=.48\textwidth]{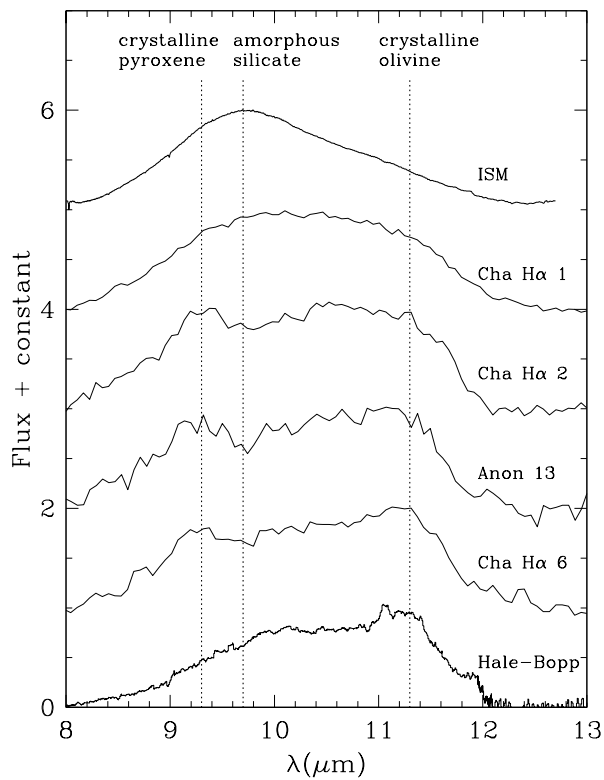}
\end{center}
\caption{\label{fig:sili}
\small Spitzer spectra of disks around low-mass stars and BDs
({\em Furlan et al.}, 2005a; {\em Apai et al.}, 2005).
The spectra have been continuum-subtracted and normalized to the peak emission
in the range between 7.6 and 13.5~\micron. 
For comparison we show the spectra of the amorphous silicate-dominated
interstellar medium and the crystalline-rich comet Hale-Bopp.}  
\end{figure}

The recent Spitzer spectroscopy of disks around low-mass stars and BDs 
has confirmed the results from the SED modeling.
Most of the spectra show a silicate emission feature broader than that from the 
interstellar medium and peaks that are indicative of crystalline grains 
(see Fig.~\ref{fig:sili}).
{\em Furlan et al.} (2005a) found that the disk model of V410~Anon~13 also 
required a reduced gas-to-dust ratio, which was suggestive of some settling.
A quantitative analysis of the dust composition of disks in
Chamaeleon~I reveals large grains and 
high crystallinity mass fractions ($\sim$40\%) for the majority of the sources
({\em Apai et al.}, 2005). In addition, most of the SEDs are consistent with 
flatter disk structures than those predicted by vertical hydrostatic 
equilibrium.
These results demonstrate that dust processing is largely independent of 
stellar properties and mainly determined by local processes in the disk. 

\subsection{\textbf{Planet Formation}}

\begin{figure}[t]
\begin{center}
\includegraphics[width=.48\textwidth]{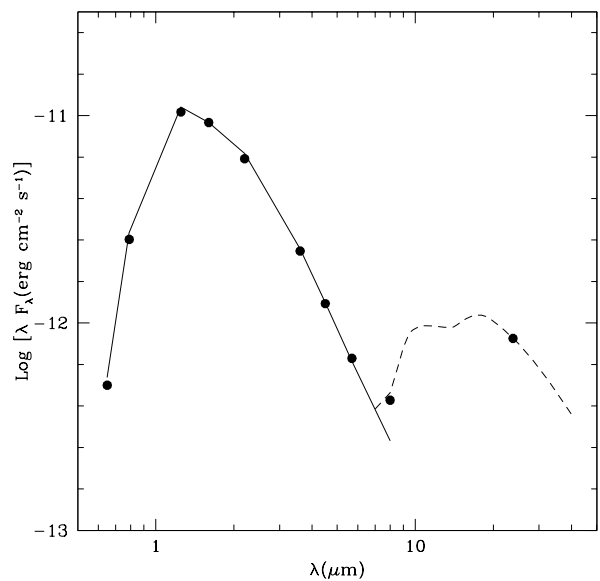}
\end{center}
\caption{\small SED of an object near the hydrogen burning limit in IC~348
({\em points, Muzerolle et al.}, 2005) compared to the
median SED of late-type members of IC 348 
that lack IR excess emission ({\em solid line, Lada et al.}, 2006).
This object exhibits excess emission only at $\lambda\geq8$~\micron, which has
been fit with a model of a disk with an inner hole ({\it dashed line}). 
\label{sed}}
\end{figure}

The identification of grain growth, crystallization, and dust settling in 
BD disks indicate that the first steps leading to planet formation 
occur in disks around BDs.
In fact, there now appears to be tantalizing evidence for planet formation
at a more advanced stage around a low-mass object in IC~348, source 316 from
{\em Luhman et al.} (2003b).
This object has a spectral type of M6.5, indicating a mass near the hydrogen 
burning mass limit, and no strong signature of accretion based on its small
H$\alpha$ equivalent width.  Spitzer photometry
has revealed strong excess emission at 24~\micron, indicating a substantial
disk (Fig.~\ref{sed}).  Interestingly, no excess emission is seen
at wavelengths shortward of 8~\micron, strongly suggesting the presence of 
an inner hole in the disk that is cleared of at least small dust grains.  
Disk models of
the SED require an inner hole size of 0.5-1 AU to fit the observations
({\em Muzerolle et al.}, 2006).  IC~348-316 is thus the first low-mass object
known to possess significant inner disk clearing akin to that seen in
higher-mass CTTSs such as CoKu Tau/4 ({\em D'Alessio et al.}, 2005).
The origin of this clearing is a matter of considerable debate.
{\em Muzerolle et al.} (2006) ruled out the photoevaporation model, in which
a photoevaporative wind generated by UV radiation from the central object
can remove material from the inner disk (e.g., {\em Clarke et al.}, 2001),
because the mass loss timescale is much longer than the age of IC~348-316 
(1-3 Myr) given plausible UV flux from an accretion shock or chromosphere.
Other possibilities include inside-out dust coagulation into
meter or kilometer-sized planetesimals, or the rapid formation of
a single planet which is preventing further accretion from the outer disk.
For the latter scenario, {\em Muzerolle et al.} (2006) estimated a plausible 
mass range of $M_p \sim 2.5-25 \; M_{\oplus}$. This type of SED analysis 
is not conclusive proof of the presence of a planetary companion, 
but nevertheless it strongly
suggest that the same steps to planet formation interpreted from observations
of disks around stars are also possible in disks around BDs.
Higher resolution data for IC~348-316 through Spitzer spectroscopy
should better constrain the nature of its inner disk hole.

\section{Summary}

We summarize the current observations of BDs that are relevant to their 
formation as follows:

\begin{enumerate}

\item
The least massive known free-floating BDs have masses of $\sim10$~$M_{\rm Jup}$.
No conclusive measurement of the minimum mass of BDs is yet available.

\item
Stars outnumber BDs at 20-80~$M_{\rm Jup}$ by a factor of $\sim5$-8 in 
star-forming regions. This ratio is consistent with data for BDs
in the solar neighborhood,
although the larger uncertainties in the field data allow for modest 
differences from star-forming regions (factor of a few).

\item
Stars and BDs share similar velocity and spatial distributions in the 
available data for star-forming regions.

\item
In the original BD desert observed at separations less than 
3~AU from solar-type primaries, BD companions are less common
than stellar companions by a factor of $\sim100$.
BDs are outnumbered by stars at larger separations as well, but 
the size of the deficiency ($\sim3$-10) is smaller than at close separations,
and is consistent with the deficiency of BDs among isolated objects ($\sim5$-8).
These data suggest that wider stellar and substellar companions 
form in the same manner as their free-floating counterparts, and that a
true BD desert for solar-type stars is restricted to small separations.

\item
For both star-forming regions and the solar neighborhood, 
binary fractions decrease continuously with mass from stars to BDs and 
most binary BDs have small separations, although a few wide systems do exist. 

\item
Accretion rates decrease continuously with mass from stars to BDs 
(as $\mdot \propto M^2$).

\item
Circumstellar disks have been found around BDs with 
masses as low as $\sim10$~$M_{\rm Jup}$.

\item
The disk fraction of BDs is similar to that of stars at ages of a few
million years.  BDs also appear to have similar disk lifetimes, although 
a definitive statement is not possible with available data.

\item
Disks around BDs exhibit a range of geometries from flat to flared, and some 
of these disks experience grain growth and settling and may develop 
inner holes, which are possible signatures of planet formation. 
All of these characteristics are also found among disks around stars.

\end{enumerate}

All of these data are consistent with a common formation mechanism for BDs
and stars. In particular, the existence of widely separated binary BDs and 
a likely isolated proto-BD ({\em Young et al.}, 2004; {\em Bourke et al.},
2005; {\em Huard et al.}, 2006) indicate that 
some BDs are able to form in the same manner as stars through unperturbed cloud 
fragmentation.  It remains possible that additional mechanisms such as 
ejection and photoevaporation influence the birth of some BDs, just
as they likely do with stars. 
However, it appears that they are not essential ingredients in making it 
possible for these small bodies to form.

\bigskip

\centerline\textbf{ REFERENCES}
\bigskip
\parskip=0pt
{\small
\baselineskip=11pt

\refs Allen P. R., Koerner D. W., Reid I. N., and Trilling D. E. (2005)
\apj, {\em 625}, 385-397.

\refs Apai D., Pascucci I., Henning Th., Sterzik M. F., Klein R., et al. 
(2002) \apj, {\em 573}, L115-L117.

\refs Apai D., Pascucci I., Sterzik M. F., van der Bliek N., Bouwman J., 
Dullemond C. P., and Henning Th. (2004) \aap, {\em 426}, L53-L57.

\refs Apai D., Pascucci I., Bouwman J., Natta A., Henning Th., and 
Dullemond C. P. (2005) {\em Science}, {\em 310}, 834-836.

\refs Ardila D., Mart{\'\i}n E., and Basri G. (2000) \aj, {\em 120}, 479-487.

\refs Barrado y Navascu\'es D. and Mart{\'\i}n E. L. (2003) \aj, {\em 126}, 
2997-3006.

\refs Barrado y Navascu\'es D., Zapatero Osorio M. R., B\'ejar V. J. S., 
Rebolo R., Mart{\'\i}n E. L., Mundt R., and Bailer-Jones C. A. L.
(2001) \aap, {\em 377}, L9-L13.

\refs Barrado y Navascu\'es D., Zapatero Osorio M. R., Mart{\'\i}n E. L.,
B\'ejar V. J. S., Rebolo R., and Mundt R. (2002) \aap, {\em 393}, L85-L88.

\refs Barrado y Navascu\'es D., Mohanty S., and Jayawardhana R. (2004a) \apj,
{\em 604}, 284-296.

\refs Barrado y Navascu\'es D., Stauffer J. R., Bouvier J., Jayawardhana R.,
and Cuillandre J.-C. (2004b) \apj, {\em 610}, 1064-1078.

\refs Baraffe I., Chabrier G., Allard F., and Hauschildt P. H. (1998)
\aap, {\em 337}, 403-412.

\refs Basri G. and Mart\'{\i}n E. L. (1999) \apj, {\em 118}, 2460-2465. 

\refs Bate M. R. and Bonnell I. A. (2005) \mnras, {\em 356}, 1201-1221.

\refs Bate M. R., Bonnell I. A., and Bromm V. (2002) \mnras, {\em 332},
L65-L68.

\refs Bate M. R., Bonnell I. A., and Bromm V. (2003) \mnras, {\em 339},
577-599.

\refs B\'ejar V. J. S., Zapatero Osorio M. R., and Rebolo R. (1999) \apj,
{\em 521}, 671-681.

\refs B\'ejar V. J. S., Mart\'{\i}n E. L., Zapatero Osorio M. R., Rebolo R.,
Barrado y Navascu\'es D., et al. (2001) \apj, {\em 556}, 830-836.

\refs Bill\`{e}res M., Delfosse X., Beuzit J.-L., Forveille T., Marchal L.,
and Mart\'{\i}n E. L. (2005) \aap, {\em 440}, L55-L58.

\refs Bonnell I. A. and Davies M. B. (1998) \mnras, {\em 295}, 691-698.

\refs Boss A. (2001) \apj, {\em 551}, L167-L170.

\refs Bourke T. L., Crapsi A., Myers P. C., Evans N. J., Wilner D. J., et al. 
(2005) \apj, {\em 633}, L129-L132.

\refs Bouy H., Brandner W., Mart{\'\i}n E. L., Delfosse X., Allard F., 
and Basri G. (2003) \aj, {\em 126}, 1526-1554.

\refs Bouy H., Brandner W., Mart{\'\i}n E. L., Delfosse X., Allard F., et al. 
(2004) \aap, {\em 424}, 213-226.

\refs Bouy H., Mart{\'\i}n E. L., Brandner W., Zapatero Osorio M. R.,
B\'ejar V. J. S., et al. (2006) \aap, {\em in press}.

\refs Brice\~no C., Hartmann L., Stauffer J., and Mart\'in E. (1998) \aj,
{\em 115}, 2074-2091.

\refs Brice\~{n}o C., Luhman K. L., Hartmann L., Stauffer J. R., and
Kirkpatrick J. D. (2002) \apj, {\em 580}, 317-335.

\refs Burgasser A. J., Kirkpatrick J. D., Reid I. N., Brown M. E., Miskey C. L.,
and Gizis J. E. (2003) \apj, {\em 586}, 512-526.

\refs Burgasser A. J., Kirkpatrick J. D., McGovern M. R., McLean I. S., 
Prato L., and Reid, I. N. (2004) \apj, {\em 604}, 827-831.

\refs Burrows, A., Marley M., Hubbard W. B., Lunine J. I., Guillot T., et al.
(1997) \apj, {\em 491}, 856-875.

\refs Caballero J. A., B\'ejar V. J. S., Rebolo R., and Zapatero Osorio M. R. 
(2004) \aap, {\em 424}, 857-872.

\refs Calvet N. and Gullbring E. (1998) \apj, {\em 509}, 802-818.

\refs Chabrier G. (2002) \apj, {\em 567}, 304-313.

\refs Chabrier G., Baraffe I. Allard F., and Hauschildt P. H. (2000) 
\apj, {\em 542}, 464-472.

\refs Chauvin G., Lagrange A.-M., Dumas C., Zuckerman B., Mouillet, D., et al.
(2004) \aap, {\em 425}, L29-L32.

\refs Chauvin G., Lagrange A.-M., Dumas C., Zuckerman B., Mouillet, D., et al. 
(2005) \aap, {\em 438}, L25-L28.

\refs Chiang E. I. and Goldreich P. (1997) \apj, {\em 490}, 368-376.

\refs Chiang E. I. and Goldreich P. (1999) \apj, {\em 519}, 279-284.

\refs Clarke C. J., Gendrin A., and Sotomayor M. (2001) \mnras, {\em 328},
485-491.

\refs Close L. M., Siegler N., Freed M., and Biller B. (2003) \apj, 
{\em 587}, 407-422.

\refs Comer\'{o}n F., Rieke G. H., Claes P., Torra J., and Laureijs R. J.
(1998) \aap, {\em 335}, 522-532.

\refs Comer\'{o}n F., Rieke G. H., and Neuh\"auser R. (1999) \aap, {\em 343},
477-495.

\refs Comer\'{o}n F., Neuh\"auser R., and Kaas A. A. (2000) \aap, {\em 359},
269-288.

\refs Comer\'{o}n F., Reipurth B., Henry A., and Fern\'{a}ndez M. (2004) 
\aap, {\em 417}, 583-596.

\refs Cushing M. C., Tokunaga A. T., and Kobayashi N. (2000) \aj, {\em 119},
3019-3025.

\refs D'Alessio P., Hartmann L., Calvet N., Franco-Hern\'andez R., 
Forrest W. J., et al. (2005) \apj, {\em 621}, 461-472.

\refs Delgado-Donate E. J., Clarke C. J., and Bate M. R. (2004) \mnras, 
{\em 347}, 759-770.

\refs Dobashi, K., Uehara, H., Kandori, R., Sakurai, T., Kaiden, M., et al.
(2005) \pasj, {\em 57}, S1-S386.

\refs Duch\^ene G., Bouvier J., and Simon T. (1999) \aap, {\em 343}, 831-840.

\refs Dullemond C. P. and Dominik C. (2005) \aap, {\em 434}, 971-986.

\refs Duquennoy A. and Mayor M. (1991) \aap, {\em 248}, 485-524.

\refs Fazio G. G., Hora J. L., Allen L. E., Ashby M. L. N., Barmby P., et al.
(2004) \apjs, {\em 154}, 10-17.

\refs Feigelson E. D., Gaffney J. A., Garmire G., Hillenbrand L. A., and 
Townsley L. (2003) \apj, {\em 584}, 911-930.

\refs Fern\'andez M. and Comer\'on F. (2001) \aap, {\em 380}, 264-276.

\refs Fischer D. A. and Marcy G. W. (1992) \apj, {\em 396}, 178-194.

\refs Furlan E., Calvet N., D'Alessio P., Hartmann L., Forrest W. J., et al. 
(2005a) \apj, {\em 621}, L129-L132.

\refs Furlan E., Calvet N., D'Alessio P., Hartmann L., Forrest W. J., et al. 
(2005b) \apj, {\em 628}, L65-L68.

\refs Gizis J. E. (2002) \apj, {\em 575}, 484-492.

\refs Gizis J. E., Kirkpatrick J. D., Burgasser A., Reid I. N., Monet D. G,
et al. (2001) \apj, {\em 551}, L163-L166.

\refs Gizis J. E., Reid I. N., Knapp G. R., Liebert J., Kirkpatrick J. D., 
et al. (2003) \aj, {\em 125}, 3302-3310.

\refs Gizis J. E., Shipman H. L., and Harvin J. A. (2005) \apj, {\em 630},
L89-L91.

\refs Glassgold A. E., Najita J., and Igea J. (2004) \apj, {\em 615}, 
972-990.

\refs Guenther E. W. and Wuchterl G. (2003) \aap, {\em 401}, 677-683.

\refs Guieu S., Dougados C., Monin J.-L., Magnier E., and Martin E. L.
(2006), \aap, {\em 446}, 485-500.

\refs Gullbring E., Hartmann L., Brice\~{n}o C., and Calvet N. (1998) \apj,
{\em 492}, 323-341.

\refs Haisch K. E., Lada E. A., and Lada C. J. (2001) \apj, {\em 553}, 
L153-L156.

\refs Harrington R. S., Dahn C. C., and Guetter H. H. (1974) \apj, 
{\em 194}, L87-L87.

\refs Hartmann L., Megeath S. T., Allen L. E., Luhman K. L., Calvet N., et al.
(2005) \apj, {\em 629}, 881-896.

\refs Henning Th., Dullemond C. P., Wolf S., and Dominik C. (2006) 
In {\em Planet Formation. Theory, Observation and Experiments}
(H. Klahr and W. Brandner, eds.), Cambridge Univ. {\em in press}.

\refs Hillenbrand L. A. (1997) \aj, {\em 113}, 1733-1768.

\refs Hillenbrand L. A. and Carpenter J. M. (2000) \apj, {\em 540}, 236-254.

\refs Hillenbrand L. A., Strom S. E., Calvet N., Merrill K. M., Gatley I., 
et al. (1998) \aj, {\em 116}, 1816-1841.

\refs Huard T. L., Myers P. C., Murphy D. C., Crews L. J., Lada C. J., et al. 
(2006) \apj, {\em in press}.

\refs Jayawardhana R., Ardila D. R., Stelzer B., and Haisch K. E. (2003a)
\aj, {\em 126}, 1515-1521.

\refs Jayawardhana R., Mohanty S., and Basri G. (2003b) \apj, {\em 592}, 
282-287.

\refs Joergens V. (2006a) \aap, {\em 446}, 1165-1176.

\refs Joergens V. (2006b) \aap, {\em 448}, 655-663.

\refs Joergens V. and Guenther E. (2001) \aap, {\em 379}, L9-L12.

\refs Kenyon S. J. and Hartmann L. (1987) \apj, {\em 323}, 714-733.

\refs Kenyon S. J. and Hartmann L. (1995) \apjs, {\em 101}, 117-171.

\refs Kirkpatrick J. D., Barman T. S., Burgasser A. J., McGovern M. R., 
McLean I. S., et al. (2006) \apj, {\em in press}.

\refs Klein R., Apai D., Pascucci I., Henning Th., and Waters L. B. F. M.
(2003) \apj, {\em 593}, L57-L60.

\refs Kraus A. L., White R. J., and Hillenbrand L. A. (2005) \apj, 
{\em 633}, 452-459.

\refs Kraus A. L., White R. J., and Hillenbrand L. A. (2006) \apj, 
{\em in press}.

\refs Kroupa P. and Bouvier J. (2003) \mnras, {\em 346}, 369-380.

\refs Lada C. J., Muench A. A., Haisch K. E., Lada E. A., Alves J. F., et al.
(2000) \aj, {\em 120}, 3162-3176.

\refs Lada C. J., Muench A. A., Lada E. A., and Alves J. F. (2004) \aj,
{\em 128}, 1254-1264.

\refs Lada C. J., Muench A. A., Luhman K. L., Allen L., Hartmann L., et al.
(2006) \aj, {\em in press}.

\refs Levine J. L., Steinhauer A., Elston R. J., and Lada E. A. (2006) \apj, 
{\em submitted}.

\refs Liu M. C., Najita J., and Tokunaga A. T. (2003) \apj, {\em 585}, 372-391.

\refs Lucas P. W., Roche P. F., Allard F., and Hauschildt P. H. (2001)
\mnras, {\em 326}, 695-721.

\refs Lucas P. W., Roche P. F., and Tamura M. (2005) \mnras, {\em 361}, 211-232.

\refs Luhman K. L. (1999) \apj, {\em 525}, 466-481.

\refs Luhman K. L. (2000) \apj, {\em 544}, 1044-1055.

\refs Luhman K. L. (2004a) \apj, {\em 602}, 816-842.

\refs Luhman K. L. (2004b) \apj, {\em 614}, 398-403.

\refs Luhman K. L. (2004c) \apj, {\em 617}, 1216-1232.

\refs Luhman K. L. (2005) \apj, {\em 633}, L41-L44.

\refs Luhman K. L. (2006) \apj, {\em submitted}.

\refs Luhman K. L. and Potter D. (2006) \apj, {\em 638}, 887-896.

\refs Luhman K. L., Liebert J., and Rieke G. H. (1997) \apj, {\em 489}, 
L165-L168.

\refs Luhman K. L., Rieke G. H., Lada C. J., and Lada E. A. (1998)
\apj, {\em 508}, 347-369.

\refs Luhman K. L., Rieke G. H., Young E. T., Cotera A. S., Chen H., et al.
(2000) \apj, {\em 540}, 1016-1040.

\refs Luhman K. L., Brice\~{n}o C., Stauffer J. R., Hartmann L., 
Barrado y Navascu\'es D., and Caldwell N. (2003a) \apj, {\em 590}, 348-356.

\refs Luhman K. L., Stauffer J. R., Muench A. A., Rieke G. H., Lada E. A., 
et al. (2003b) \apj, {\em 593}, 1093-1115.

\refs Luhman K. L., Peterson D. E., and Megeath S. T. (2004) \apj, {\em 617},
565-568.

\refs Luhman K. L., Lada E. A., Muench A. A., and Elston R. J. (2005a) \apj, 
{\em 618}, 810-816.

\refs Luhman K. L., D'Alessio P., Calvet N., Allen L. E., Hartmann L., et al.
(2005b) \apj, {\em 620}, L51-L54.

\refs Luhman K. L., McLeod K. K., and Goldenson N. (2005c) \apj, {\em 623}, 
1141-1156.

\refs Luhman K. L., Lada C. J., Hartmann L., Muench A. A., Megeath S. T., 
et al. (2005d) \apj, {\em 631}, L69-L72.

\refs Luhman K. L., Adame L., D'Alessio P., Calvet N., Hartmann L., et al.
(2005e) \apj, {\em 635}, L93-L96.

\refs Marcy G. W. and Butler R. P. (2000) \pasp, {\em 112}, 137-140.

\refs Marcy G., Butler R. P., Fischer D., Wright J. T., Tinney C. G., and
Jones H. R. A. (2005) {\em Progress of Theoretical Physics Supplement}, 
{\em 158}, 24-42.

\refs Mart{\'\i}n E. L. and Zapatero Osorio M. R. (2003) \apj, 593, L113-L116.

\refs Mart{\'\i}n E. L., Brander W., Bouvier J., Luhman K. L., Stauffer J., 
et al. (2000) \apj, {\em 543}, 299-312.

\refs Mart{\'\i}n E. L., Zapatero Osorio M. R., Barrado y Navascu\'es D.,
B\'ejar V. J. S., Rebolo R., et al. (2001a) \apj, {\em 558}, L117-L121.

\refs Mart{\'\i}n E. L., Dougados C., Magnier E., M\'enard F., Magazz\`u A., 
et al. (2001b) \apj, {\em 561}, L195-L198.

\refs Mart{\'\i}n E. L., Delfosse X., and Guieu S. (2004) \aj, {\em 127}, 
449-454.

\refs Masciadri E. and Raga A. C. (2004) \apj, {\em 615}, 850-854.

\refs Mazeh T., Goldberg D., Duquennoy, A., and Mayor M. (1992) \apj, {\em 265},
265-268.

\refs McCarthy C. and Zuckerman B. (2004) \aj, {\em 127}, 2871-2884.

\refs McGehee P. M., West A. A., Smith J. A., Anderson K. S. J., and 
Brinkmann J. (2005) \aj, {\em 130}, 1752-1762.

\refs Mohanty S. and Basri G. (2003) \apj, {\em 583}, 451-472.

\refs Mohanty S., Jayawardhana R., and Barrado y Navascu\'{e}s D. (2003)
\apj, {\em 593}, L109-L112.

\refs Mohanty S., Jayawardhana R., Natta A., Fujiyoshi T., Tamura M., and 
Barrado y Navascu\'{e}s D. (2004) \apj, {\em 609}, L33-L36.

\refs Mohanty S., Jayawardhana R., and Basri G. (2005) \apj, {\em 626}, 
498-522.

\refs Moraux E. and Clarke C. (2005) \aap, {\em 429}, 895-901. 

\refs Muench A. A., Alves J., Lada C. J., and Lada E. A. (2001) \apj, 
{\em 558}, L51-L54.

\refs Muench A. A., Lada E. A., Lada C. J., and Alves J. (2002) \apj, 
{\em 573}, 366-393.

\refs Muench A. A., Lada E. A., Lada C. J., Elston R. J., Alves J. F., et al.
(2003) \aj, {\em 125}, 2029-2049.

\refs Muzerolle J., Hartmann L., and Calvet N. (1998) \aj, {\em 116}, 455-468.

\refs Muzerolle J., Brice\~no C., Calvet N., Hartmann L., Hillenbrand L., 
and Gullbring E. (2000) \apj, {\em 545}, L141-L144.

\refs Muzerolle J., Calvet N., and Hartmann L. (2001) \apj, {\em 550}, 944-961.

\refs Muzerolle J., Hillenbrand L., Calvet N., Brice\~no C., and Hartmann L. 
(2003) \apj, {\em 592}, 266-281.

\refs Muzerolle J., Luhman K. L., Brice\~{n}o C., Hartmann L., and Calvet N. 
(2005) \apj, {\em 625}, 906-912.

\refs Muzerolle J., Adame L., D'Alessio P., Calvet N., Luhman K. L., et al.
(2006) \apj, {\em in press}.

\refs Natta A. and Testi L. (2001) \aap, {\em 376}, L22-L25

\refs Natta A., Testi L., Comer\'on F., Oliva E., D'Antona F., et al. (2002) 
\aap, {\em 393}, 597-609.

\refs Natta A., Testi L., Muzerolle J., Randich S., Comer\'on F., and Persi P.
(2004) \aap, {\em 424}, 603-612.

\refs Neuh\"auser R. and Comer\'{o}n F. (1999) \aap, {\em 350}, 612-616.

\refs Neuh\"auser R., Brandner W., Alves J., Joergens V. and Comer\'on F. 
(2002) \aap, {\em 384}, 999-1011.

\refs Oasa Y., Tamura M., and Sugitani K. (1999) \apj, {\em 526}, 336-343.

\refs Padoan P. and Nordlund A. (2004) \apj, {\em 617}, 559-564.

\refs Padoan P., Kritsuk A., Norman M. L., and Nordlund A. (2005) \apj, 
{\em 622}, L61-L64.

\refs Pascucci I., Apai D., Henning Th., and Dullemond C. P. (2003) \apj,
{\em 590}, L111-L114.

\refs Persi P., Marenzi A. R., Olofsson G., Kaas A. A., Nordh L., et al. (2000) 
\aap, {\em 357}, 219-224.

\refs Phan-Bao N., Mart{\'\i}n E. L., Reyl\'e C., Forveille T., and Lim J.
(2005) \aap, {\em 439}, L19-L22.

\refs Przygodda F., van Boekel R., \`Abrah\`am P., Melnikov S. Y., Waters L. B.
F. M., Leinert Ch. (2003) \aap, {\em 412}, L43-L46.

\refs Reid I. N., Gizis J. E., Kirkpatrick J. D., and Koerner D. W. (2001) \aj, 
{\em 121}, 489-502.

\refs Reid I. N., Kirkpatrick J. D., Liebert J., Gizis J. E., Dahn C. C., and 
Monet D. G. (2002) \aj, {\em 124}, 519-540.

\refs Reipurth B. and Clarke C. (2001) \aj, {\em 122}, 432-439.

\refs Scholz A. and Eisl\"offel J. (2004) \aap, {\em 419}, 249-267.

\refs Scholz A. and Eisl\"offel J. (2005) \aap, {\em 429}, 1007-1023.

\refs Scholz A., Jayawardhana R., and Brandeker A. (2005) \apj, {\em 629},
L41-L44.

\refs Scholz R.-D., McCaughrean M. J., Zinnecker H., and Lodieu N. (2005) \aap, 
{\em 430}, L49-L52.

\refs Siegler N., Close L. M., Cruz K. L., Mart{\'\i}n E. L., and Reid I. N. 
(2005) \apj, {\em 621}, 1023-1032.

\refs Slesnick C. L., Hillenbrand L. A., and Carpenter J. M. (2004) \apj, 
{\em 610}, 1045-1063.

\refs Stassun K., Mathieu R. D., Vaz L. P. V., Valenti J. A., and Gomez Y.
(2006) {\em Nature, in press}.

\refs Stauffer J. R., Hamilton D., and Probst R. (1994) \aj, {\em 108}, 
155-159.

\refs Sterzik M. F. and Durisen R. H. (2003) \aap, {\em 400}, 1031-1042.

\refs Sterzik M. F., Pascucci I., Apai D., van der Bliek N., and Dullemond 
C. P. (2004) \aap, {\em 427}, 245-250.

\refs Testi L., Natta A., Oliva E., D'Antona F., Comeron F. et al. (2002) \apj, 
{\em 571}, L155-L159.

\refs Umbreit S., Burkert A., Henning Th., Mikkola S., and Spurzem R. (2005) 
\apj, {\em 623}, 940-951.

\refs van Boekel R., Min M., Waters L. B. F. M., de Koter A., Dominik C., 
et al. (2005) \aap, {\em 437}, 189-208.

\refs Walker C., Wood K., Lada C. J., Robitaille T., Bjorkman J. E., and 
Whitney B. (2004) \mnras, {\em 351}, 607-616.

\refs Werner M. W., Roellig T. L., Low F. J., Rieke G. H., Rieke M., et al.
(2004) \apjs, {\em 154}, 1-9.

\refs Whelan E. T., Ray T. P., Bacciotti F., Natta A., Testi, L. and Randich S. 
(2005) {\em Nature}, {\em 435}, 652-654.

\refs White R. J. and Basri G. (2003) \apj, {\em 582}, 1109-1122.

\refs White R. J. and Ghez A. M. (2001) \apj, {\em 556}, 265-295.

\refs White R. J. and Hillenbrand L. A. (2004) \apj, {\em 616}, 998-1032.

\refs White R. J., Ghez A. M., Reid I. N., and Schultz G. (1999) \apj, 
{\em 520}, 811-821.

\refs Whitworth A. P. and Zinnecker H. (2004) \aap, {\em 427}, 299-306.

\refs Wilking B. A., Greene T. P., and Meyer M. R. (1999) \aj, {\em 117}, 
469-482.

\refs Wilking B. A., Meyer M. R., Greene T. P., Mikhail A., and Carlson G.
(2004) \aj, {\em 127}, 1131-1146.

\refs Zapatero Osorio M. R., B\'ejar V. J. S., Rebolo R., Mart{\'\i}n E. L.,
and Basri G. (1999) \apj, {\em 524}, L115-L118.

\refs Zapatero Osorio M. R., B\'ejar V. J. S., Mart{\'\i}n E. L., Rebolo R., 
Barrado y Navascu\'es D., et al. (2000) {\em Science}, {\em 290}, 103-107.

\refs Zapatero Osorio, M. R., B\'ejar V. J. S., Mart{\'\i}n E. L.,
Barrado y Navascu\'es D., and Rebolo R. (2002a) \apj, {\em 569}, L99-L102.

\refs Zapatero Osorio, M. R., B\'ejar V. J. S., Mart{\'\i}n E. L., Rebolo R.,
Barrado y Navascu\'es D., et al. (2002b) \apj, {\em 578}, 536-542.

\refs Zapatero Osorio, M. R., B\'ejar V. J. S., Pavlenko Y., Rebolo R., 
Allende P., et al. (2002c) \aap, {\em 384}, 937-953.

\refs Young, C. H., J{\o}rgensen J. K., Shirley Y. L., Kauffmann J., Huard T.,
et al. (2004) \apjs, {\em 154}, 396-401.

\end{document}